\documentclass[%
 amsmath,amssymb,
 aps,
 prx,twocolumn,english,superscriptaddress,floatfix,longbibliography
]{revtex4-2}

\usepackage{algorithm}
\usepackage{algpseudocode}

\usepackage{xr}
\makeatletter
\newcommand*{\addFileDependency}[1]{
  \typeout{(#1)}
  \@addtofilelist{#1}
  \IfFileExists{#1}{}{\typeout{No file #1.}}
}
\makeatother

\newcommand*{\myexternaldocument}[1]{
    \externaldocument{#1}
    \addFileDependency{#1.tex}
    \addFileDependency{#1.aux}
}

\newcommand\foldername{.}

\myexternaldocument{\foldername/SI}

\usepackage{graphicx}
\usepackage{dcolumn}
\usepackage{bm}
\usepackage{enumitem}
\usepackage[caption=false]{subfig}
\usepackage{dsfont}
\usepackage{accents}
\usepackage{xstring}
\usepackage{hyperref}
\usepackage{cleveref}

\usepackage{relsize}

\usepackage{tikz}
\newenvironment{captivy}[1]{
  \begin{tikzpicture}[every node/.style={inner sep=0}]
    \node[anchor=south west,inner sep=0] (image) at (0,0) {#1};
    \begin{scope}[x={(image.south east)},y={(image.north west)}]
}%
{
        \end{scope}%
  \pgfresetboundingbox
  \path[use as bounding box] (image.south west) rectangle (image.north east);
  \end{tikzpicture}%
}

\newcommand*{\oversubcaption}[3]{%
  \draw (#1) node[fill=white,inner sep=0pt, opacity=0.2, above, yscale=1.1, xscale=1.1] {\phantom{(a)#2}};
  \draw (#1) node[inner sep=0pt, above]{%
    \subfloat[#2\label{#3}]{\phantom{(a)}}
  };
}

\newcommand{\stripFigNum}[1]{%
  \StrLen{\arabic{figure}}[\figlen]
  \StrGobbleLeft{#1}{\figlen}
}

\Crefname{figure}{Fig.}{Figs.}
\Crefname{equation}{Eq.}{Eqs.}
\Crefname{section}{Sec.}{Secs.}
\creflabelformat{equation}{#2\textup{#1}#3}

\newcommand{\Crefsubfigrange}[3]{%
  \Crefrange{#1#2}{#1#3}%
}
\crefrangelabelformat{figure}{#3#1#4-#5\stripFigNum{#2}#6}

\newcommand{\Crefsubfiglist}[3]{%
  \Cref{#1#2},\hyperref[#1#3]{#3}%
}

\begin{document}

\preprint{APS/123-QED}

\title{Slow Relaxation and Landscape-Driven Dynamics in Viscous Ripening Foams}

\author{Amruthesh Thirumalaiswamy}
\author{Clary Rodr{\'i}guez-Cruz}
\author{Robert A. Riggleman}
 \email{rrig@seas.upenn.edu}
\author{John C. Crocker}%
 \email{jcrocker@seas.upenn.edu}
\affiliation{%
 Department of Chemical and Biomolecular Engineering, University of Pennsylvania, Philadelphia, Pennsylvania\\
}%

\begin{abstract}
Foams and dense emulsions display complex mechanical behavior, including intermittent rearrangement dynamics, power-law rheology, and slow recovery after perturbation. These effects have long been considered evidence for glassy physics in these and other materials having similar mechanics, such as the cytoskeleton. Here we study such anomalous mechanics in a simulated wet foam driven by ripening and find behavior that has a different physical origin than that in glasses. Rather, the dynamics is due to a balance of forces from the system's self-similar potential energy landscape and viscous stress. At the lowest viscosities, bubbles move intermittently, with the system shifting abruptly between shallow potential energy minima. For higher viscosities, in contrast, the bubbles move continuously and the system follows a tortuous, fractal path through high-dimensional configuration space, at higher mean energy than the lower viscosity case. The long-time dynamics and power-law rheology are the direct consequence of the potential energy landscape's self-similar geometry. Lastly, we find that the slow recovery of perturbed foams is due to the foam being kinetically rather than energetically trapped in high-energy portions of the energy landscape. Overall, viscous ripening foams follow a biased energy minimization pathway that explores regions of the energy landscape that are qualitatively different (flatter and smoother) than those corresponding to well-annealed glasses.
\end{abstract}


\maketitle

Multiphase complex fluids—such as foams and dense emulsions—often exhibit unexpected and complex physical \cite{FoamsBookWeaire1999, SGMReviewJoshi2014} and rheological \cite{RheologyreviewHohler2005, FoamrheologyreviewDenkov2009, RheologyFoamsReviewDollet2014} behavior, particularly power-law rheology and super-diffusion \cite{FlowBlair2016,MultipleGiavazzi2020,ViscoelasticVeronique2022, FoamsClary2022, AnomalousGuidolin2024} and slow recovery following mechanical perturbations \cite{PerturbfoamsGopal1995, MemoryFoamHohler1999, PerturbationsFoamHohler2001}. Similar “glassy” mechanics have been observed in cytoskeletal networks \cite{SGRCellsFabry2001, CytoSkeletalFredberg2005, CellMechCrocker2006, CytoSGRMandadapu2008, CellMechCrocker2009} and granular materials \cite{granularKou2017}. Early models of dry foams focused on mechanically stable polyhedral bubble structures \cite{PottsGrest1988,VertexBrakke1992,SurfEvolOkuzono1993}, while wet foams were typically represented as spherical bubbles with repulsive harmonic interactions \cite{FoamMechDurian1995, BubbleDurian1997}. The unusual dynamics of these athermal, jammed materials was hypothesized to be analogous to those of glasses  \cite{JammingLiuNagel1998} and due to the slow hopping of the system over high-energy barriers between stable minima  \cite{PerturbfoamsGopal1995, MemoryFoamHohler1999, PerturbationsFoamHohler2001}, culminating in the soft glassy rheology (SGR) model \cite{SGMCates1997, SGMSollich1998, MCTSGMLequex1998, SGRSollich2006}.
A pivotal 2012 study \cite{GlassJamSollich2012}, however, showed that in these systems the glass and jamming transitions are qualitatively distinct, with the mechanics of athermal foams driven by steady shear being dictated by jamming rather than glass physics. More recent simulations \cite{SGMHwang2016} and experiments \cite{FoamsClary2022} suggest that the slow dynamics of ripening foams arise not from glass-like activated hopping, but from an energy minimizing pathway constrained by a self-similar potential energy landscape, 
termed\textit{fractal landscape dynamics} (FLD).

This study aims to disentangle the respective roles of energy landscape structure, viscous stress, ripening dynamics, and glassy physics in producing the varied behaviors reported in the literature \cite{FlowBlair2016,ViscoelasticVeronique2022, FoamsClary2022,PerturbfoamsGopal1995, MemoryFoamHohler1999, PerturbationsFoamHohler2001, duri2009resolving, Sessoms2010unexpected}. We use an athermal bubble model \cite{FoamMechDurian1995, BubbleDurian1997, FoamTewari2003, GlassJamSollich2012}, driven solely by bubble radii evolving via simulated Ostwald ripening \cite{FoamStevenson2010}. The system's dynamics are controlled by the ratio of viscous damping to the ripening rate, defining a Deborah number (\textit{De}). At low \textit{De}, the system exhibits an intermittent and superdiffusive dynamics consistent with an earlier study using an infinitesimal viscosity \cite{SGMHwang2016}. At higher \textit{De}, the system evolves continuously at elevated average energy without encountering any energy minima. Across a wide range of intermediate \textit{De}, the long-time dynamics and power-law rheology are governed by FLD (i.e. are landscape-driven), while the short-time behavior is ballistic and elastic.

Further analysis reveals four key findings. First, the self-similarity leading to FLD is a generic feature of the energy landscape, even at high energies, and is present even when ripening is turned off.  Second, the short-time ballistic motion depends on viscous stress (or \textit{De}) rather than landscape geometry. Third, ripening foams at very low \textit{De} sample low energy minima similar to supercooled fluid and glassy states but not deep glass configurations. Fourth, the slow recovery of perturbed foams \cite{PerturbfoamsGopal1995, MemoryFoamHohler1999, PerturbationsFoamHohler2001} is due to kinetic trapping: the system moves slowly through a labyrinthine high-energy landscape, limited by viscous drag rather than energetic barriers.

Overall, when combined with an earlier study  \cite{MIMSEThirumalaiswamy2022}, our findings here can be explained by a `multifractal' energy landscape where high and low energy domains have different self-similar geometry. Ripening (and presumably shear) driven viscous systems evolve according to energy minimization, but biased in a direction dictated by mechanical strain, resulting in them exploring a high energy domain that is relatively smooth and flat. Very slowly deformed foams explore the region near the boundary between the high energy (foamy) and low energy (glassy) domains, filled with energy minima corresponding to the inherent structures of super-cooled fluids and glasses. Thermalized low-temperature soft-sphere systems \cite{GlassJamSollich2012,MIMSEThirumalaiswamy2022}, in contrast, evolve according to unbiased energy minimization and fall into the lower energy domain, corresponding to deep glass configurations and their associated slow dynamics. This physical picture appears highly analogous to deep learning models, where different optimizers \cite{DeepLearningKeskar2017} lead to convergence in either more expressive, flatter regions or less expressive, glassy regions of a multifractal loss landscape \cite{OptimizationLy2025}.

\section{\label{sec:bubblemodel}Ripening Bubble Models and Deborah Number \textit{De}}
We model a ripening wet foam using the bubble model \cite{FoamMechDurian1995, BubbleDurian1997} with a simplified damping rule and simulated Ostwald ripening. While the bubble model has traditionally been used to simulate foams \cite{FoamMechDurian1995, FoamTewari1999}, it also serves as an effective model for many other soft, jammed, and glassy systems \cite{RCPLiu2002, GlassJamSollich2012, DPDWarren2017}. The bubbles in this model are treated as athermal, non-inertial soft-sphere particles that interact via a pairwise repulsive harmonic potential, described in \hyperref[sec:Methods]{Methods}. The positions of bubbles are evolved using an (overdamped) equation of motion \cite{FoamMechDurian1995}. Notably, we consider a simplified version of the viscous force, $\mathbf{F}_i = \xi \mathbf{v}_i$ \cite{BubbleDurian1997, GlassJamSollich2012}, on each bubble to reduce computational effort while preserving relevant model physics,
\begin{equation}
\begin{split}
    \xi \frac{d\mathbf{r}_i}{dt} = -\mathlarger{\mathlarger{\sum}}_{j}^{nn} \frac{\partial V(r_{ij})}{\partial r_i}
\end{split}
\label{eq:damped-equation}
\end{equation}
with the right-hand side representing a summation over neighboring bubbles that contribute to the force on bubble $i$. The left side is the viscous drag force, $\xi$ being the effective viscous damping factor. 

To simulate mass exchange between bubbles due to Ostwald ripening \cite{FoamStevenson2010}, the bubble radii $\{a_i\}$ are allowed to evolve while keeping the total volume of bubbles constant (preserving nominal total mass). Ripening causes larger bubbles to grow and smaller ones to shrink over time via a diffusive mass flux, primarily between contacting bubbles. We modeled this process using the approach used in a previous study \cite{SGMHwang2016}, with a mass flux $Q$ computed using: 
\begin{equation}
\begin{split}
    \frac{Q_i}{\rho} = - \alpha \bigg[ \underbrace{ \sum_{j}^{nn}\left( \frac{1}{a_i}- \frac{1}{a_j} \right) A_{\text{overlap}}}_{\text{nearest-neighbor}} \\
    \underbrace{ + b_{mf} \left( \frac{1}{a_i} - \frac{1}{\left< a \right>} \right) a_i}_{\text{mean-field}} \bigg]
\end{split}
\label{eq:mass-flux}
\end{equation}
\noindent where the first term represents the pairwise mass flux between neighboring bubbles, the second term represents a small mean-field contribution, $\rho$ is the mass density, $\alpha$ controls the rate of ripening and $b_{mf} \ll 1$ is a parameter that ensures that the smallest bubbles shrink to zero size after becoming `rattlers'. For further details see \hyperref[sec:Methods]{Methods}. 

To evolve a system of $N \approx 1000$ bubbles at a volume fraction of $\phi=0.75$ (just above its jamming volume fraction \cite{SGMHwang2016, JammingLiuNagel1998}), we integrate the equation of motion (\Cref{eq:damped-equation}) and increment the bubble radii according to (\Cref{eq:mass-flux}) at each time step, using a fixed time step that is small enough to ensure accuracy; further details in \Cref{subsecSI:integration}. This study explored a range of different damping factors, $\xi$, while keeping $\phi$ fixed just above jamming ($\phi_J \approx 0.72$). 

Our ripening simulations eventually reach a state called \textit{dynamical scaling} \cite{DyanmicscalingMullins1986, SGMHwang2016} in which the mean bubble size grows with time, while the bubble size distribution maintains the same shape ($P(a_i / \langle a \rangle)$ remains a constant), resembling experiments \cite{FoamFeitosa2006}. It should be noted that different $\xi$ simulations reach slightly different steady-state radii distributions (see \Cref{subsecSI:a2t} for details). To reduce the time taken to reach dynamical scaling, we initialize the system with a Weibull distribution of bubble radii, $P(a) \sim (k/ \lambda) (a/ \lambda)^{k-1}$, where $k = 1.75, \lambda = 0.73$, resembling that reached under quasistatic ripening, $\xi \to 0$ \cite{SGMHwang2016}. These bubbles are then randomly placed inside in a simulation box and relaxed to their nearest minima using FIRE to provide a starting configuration. For the sake of consistency and comparison, we maintain the same initial bubble radii i.e $\{a_i\}$ for all $\xi$ values; while acknowledging that higher $\xi$ values take a longer time to reach dynamical scaling, with a different steady state radii distribution (see \Cref{secSI:avalanches} for details). To construct an ensemble for these simulations, given the deterministic nature of \Cref{eq:damped-equation}, we use different random positional initializations. Lastly, we necessarily end our simulations when two bubbles grow large enough to span the periodic boundaries to form multiple contacts with each other, which occurs earlier in larger $\xi$ systems.  

Given that our model only contains two parameters with dimensions of time, $\xi$ and $\alpha$, its dynamics must be solely controlled by a single dimensionless group containing their ratio. Specifically, we can define a \textit{Deborah number} \textit{De}, expressed as the ratio of time scales associated with ripening and viscous relaxation. The viscous relaxation time from \Cref{eq:damped-equation} is $\tau_{\xi} = \xi {\langle a \rangle}^2/ \epsilon$ and the timescale associated with ripening \Cref{eq:mass-flux} is $\tau_{\alpha} = {\langle a \rangle}^2/ \alpha$ when $b_{mf} \ll 1$ as we employ here. Taking the ratio yields a \textit{ripening Deborah number}, $De = \xi \alpha / \epsilon$. See \Cref{subsecSI:Deborahnumber} for details. We vary \textit{De} over the range $\sim 10^{-6}-10^{-2}$ by scanning values of $\xi$, holding the ripening parameters constant, to capture the various dynamical regimes of our model. 

\begin{figure*}[t]
\centering
\begin{captivy}{\includegraphics[width=\linewidth]{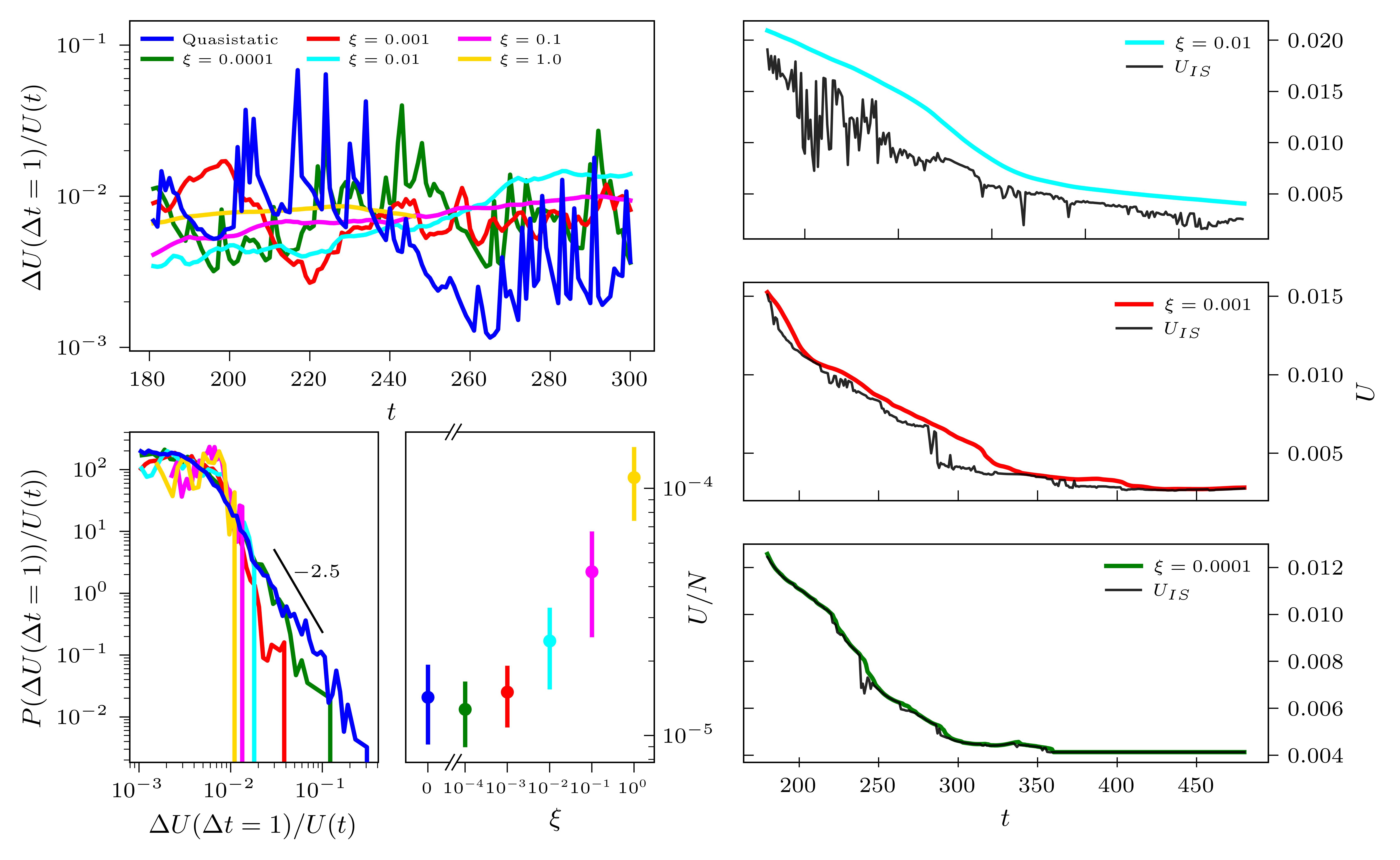}}
    \oversubcaption{0.05, 0.98}{}{fig:avalanchesa}
    \oversubcaption{0.05, 0.50}{}{fig:avalanchesb}
    \oversubcaption{0.49, 0.50}{}{fig:avalanchesc}
    \oversubcaption{0.98, 0.98}{}{fig:avalanchesd}
    \oversubcaption{0.98, 0.68}{}{fig:avalanchese}
    \oversubcaption{0.98, 0.37}{}{fig:avalanchesf}
\end{captivy}
\caption{\textbf{Intermittency in energy emerges as a feature of motion on the landscape} 
(\textbf{a}) Traces of relative energy drops $\Delta U/{U(t)}$ (for $\Delta t = 1$ or simulation points spaced by $1$ time unit) are sensitive to intermittent dynamics. For lower damping, $\xi \lesssim 0.001$, the relative change in energy shows abrupt peaks characteristic of intermittent motion and system being highly influenced by the underlying minima (shown in (d-f) above). Higher $\xi$ values show smoother energetic trends. 
(\textbf{b}) Lower $\xi$ simulations show a heavy-tailed probability distribution of energy fluctuations, typical of an avalanchey system. As the system becomes more viscous, the energy fluctuations become more Gaussian. It must be noted that any positive fluctuations or increases in system energy are ignored in our analyses. All the above analyses show the results of a single characteristic simulation run at different $\xi$.
(\textbf{c}) The average system potential energy $U$ in steady state averaged over $4$ simulations at ${\langle a \rangle}^2 \sim 0.49$ shows a clear increasing trend at higher $\xi$ values. In particular, lower $\xi$ simulations produce intermittent behavior, while for $\xi \gtrsim 0.001$, the system shifts to non-intermittent behavior.
(\textbf{d-f}) The motion of the system over the landscape is analyzed by comparing the energies traversed and corresponding minima (or IS) energies below these landscape trajectories. Periodic FIRE quenches reveal that higher $\xi$ simulations move further away from the rugged underlying landscape.}
\label{fig:avalanches}
\end{figure*}

We also performed two other simulations to compare with the above model. The first was a model with \textit{quasistatic} ripening (QR) \cite{SGMHwang2016}, which instead of integrating \Cref{eq:damped-equation} relaxes the system to the nearest local minimum of energy after every ripening step, using the FIRE algorithm \cite{FIREGumbsch2006}. The results of this model will be considered that of the $\xi = De = 0$ case. In practice, this QR model is identical to that we reported in an earlier study \cite{SGMHwang2016}, but with higher $N$ and better statistics, by virtue of an adaptive neighbor list optimized for polydisperse systems (which performs better than earlier algorithms \cite{AlgosFrenkel1996, AlgoPlimpton1995, AlgoTorquato2005}, see \Cref{secSI:algorithm} for details). The second is a \textit{random quench} (RQ) model that randomly positions bubbles in the simulation box, whose radii are drawn from a ripening simulation and then relaxed using FIRE (with no ripening moves) until the first stable energy minimum is reached.  Since the energy of random initial points in configuration space was typically very large due to unphysical overlaps, we discard the earliest portions of these energy relaxation pathways, which we defined as $U/U_{IS} \gtrsim 10$ with $U_{IS}$ the energy of the inherent structure.   

\section{\label{sec:avalanches} Intermittency is Modulated by Viscous Stress}

A notable feature of many wet foams is intermittent dynamics, where the system remains nearly at rest most of the time, punctuated by brief periods where many bubbles simultaneously move by small amounts. These events, sometimes called \textit{avalanches}, are associated with a large release of potential energy. We analyze the intermittency in our damped foam model by looking at the energy released in a unit time interval, $\Delta U (\Delta t = 1)/{ U(t)}$, shown in \Cref{fig:avalanchesa}. Two distinct limits are observed (\Cref{fig:avalanchesa}): low viscosity simulations ($\xi < 0.001$) produce large intermittent spikes in energy release corresponding to avalanches, while for higher viscosities ($\xi > 0.001$) the system evolves in a more continuous manner. This can be more clearly seen in the probability distribution or histograms of energy release, \Cref{fig:avalanchesb}, which show a heavy tail for low viscosities and Gaussian behavior for high ones. Notably, the lowest viscosity we study ($\xi = 10^{-4}$) resembles the intermittency of the quasistatic model, with some avalanches releasing about $10\%$ of total system potential energy \Cref{fig:avalanchesb}. This similarity suggests that our lowest simulated viscosity has essentially reached the limit of quasistatic dynamics.

Our previous studies of quasistatic ripening foams \cite{SGMHwang2016, FoamsClary2022} examined bubble dynamics as motion over a potential energy landscape spanning a 3$N$-dimensional configuration space. Because of ripening changing the bubble radii, this landscape slowly evolves. Avalanches corresponded to long traverses in configuration space, when the configuration moves along a steepest descent path from an energy minimum newly destabilized by ripening to the next stable one. The distance traversed between minima in turn displays a power-law tail due to the fractal clustering of minima along the energy relaxation pathway in configuration space, giving rise to power-law distributions for multiple avalanche measures \cite{SGMHwang2016} including energy, as in \Cref{fig:avalanchesb}.  The resulting dynamics were found to be features of the static geometry of the landscape, and not its slow evolution, suggesting that its large scale structure could be treated as effectively static. While this explains our findings in low-$De$ systems, it remains to be seen why systems at intermediate and higher $De$ have different behaviors.

To better understand the dynamics leading to variable intermittency, we compute the nearest underlying energy minimum for each simulation point. We do so by FIRE quenching the configurations explored by the viscous simulations and plotting both the system energy and the nearest minimum energy versus time, \Crefsubfigrange{fig:avalanches}{d}{f}. We find that low viscosity simulations $\xi \lesssim 0.001$ remain close to landscape minima, resembling the earlier findings for quasistatic foams. In foams at finite $\xi$, viscous stresses must displace the configuration away from stable energy minima. The observation of intermittent dynamics at low $\xi$ values indicates that for low viscous stresses, the system configuration remains near energy minima, and can still make intermittent large traverses between the neighborhood of some minima to the neighborhood of others. These `intermittent foams' exist near energy minima that resemble jammed packings (at volume fractions above $\phi_J$), and the distribution of such minima leads to their intermittent dynamics \cite{SGMHwang2016,FoamsClary2022}. 

In contrast, foams with intermediate viscous damping (higher \textit{De}) show continuous motion that appears decoupled from underlying energy minima, \Cref{fig:avalanchesd}, and evolve at a higher time-averaged potential energy, \Cref{fig:avalanchesc}. These systems evolve continuously in a dynamic balance between interaction and viscous forces. This loss of intermittency is clearly due to the effects of viscous stress which either steers the configuration path away from the minima, shifts the configuration to a region of configuration space that is effectively devoid of stable minima, or both (to be clarified in \Cref{sec:exploring}). Either way, the system simply doesn't encounter minima or their associated attractive wells of sufficient strength to even transiently arrest the system's dynamics. Such dynamics is qualitatively different than in a glass, where the configuration rattles around energy minima and is activated over energy barriers. These `landscape-driven' foams display continuous bubble motion and are not associated with jammed energy minima.  At still higher viscous damping  ($\xi \gtrsim 0.1$), foams display ballistic motion and elastic rheology (to be shown below). Due to their higher steady state polydispersity; however, these `viscous-driven' foams have higher jamming volume fraction $\phi_J > \phi = 0.75$. As a result, in the absence of the viscous stress associated with ripening they would be fluid suspensions without elasticity (see \Cref{secSI:avalanches} for more details).

Overall, the behavior we observe in our ripening system resembles that seen in soft sphere systems at volume fractions above $\phi_J$ subjected to steady shear \cite{BubbleDurian1997,FoamTewari1999, FoamTewari2003,GlassJamSollich2012}, despite having a somewhat broader bubble size distribution. In such systems one can define a shear Deborah number, $De_\gamma$, similar to the $De$ in our ripening case. Early simulation work \cite{BubbleDurian1997,FoamTewari1999, FoamTewari2003} showed that the dynamics become non-intermittent as $De_\gamma$ is increased, suggestive of landscape-driven dynamics similar to that we observe above. Later work \cite{GlassJamSollich2012} also showed that at volume fractions slightly above random close packing, $\phi_J$, that the system displays a $De_\gamma$-dependent increase in the average bubble stress, $\langle \sigma \rangle$, similar to \Cref{fig:avalanchesc}. 

\section{\label{sec:microrheology} Numerical microrheology of viscous foams}

The rheology of foams and dense emulsions is typically characterized as solid-like and weakly frequency dependent, often having a power-law form, $G^*(\omega) \sim \omega^\beta$, with $\beta$ in the range $\sim 0.1-0.2$. Experiments give varying results, with some showing this power-law scaling \cite{ViscoelasticVeronique2022, FoamsClary2022} and others displaying an elastic plateau \cite{RheologyHohler1998, FoamGopal2003}. Ripening causes the foam to continuously soften (as bubbles grow larger), which further complicates the accurate measurement of rheological exponents in experiments. While the SGR theory \cite{SGMSollich1998} predicts such power-law rheology, its theoretical assumptions of a glass-like potential energy landscape and granular activation appear at odds with later studies \cite{GlassJamSollich2012, SGMHwang2016} and our findings in the previous section. Capturing the low-frequency response to applied strains in simulation can be computationally expensive, however, making the determination of rheology difficult or impractical. 

We overcome the challenges of determining rheology from simulation data by using a microrheological approach suitable for systems with active fluctuations \cite{G*Mason2000, MicroRheoLubensky2003, SGMHwang2016}. This approach considers an active material as a continuum with a random fluctuating strain and stress that are related at every point by linear response and a viscoelastic constitutive equation.  In principle, by computing the power spectra of the fluctuating shear strain and stress, the dynamic shear modulus $G^*(\omega)$ can be computed from their ratio \cite{MicroRheoLubensky2003, 2ptMicroRheoCrocker2007, SGMHwang2016},
\begin{equation}
    {\lvert{G^*(\omega)}\rvert}^2 \simeq \frac{  \left<a\right>^2 \widetilde{\Delta \sigma^2}(\omega)}{3\pi\widetilde{\Delta\mathbf{r}^2}(\omega)},
    \label{eq:microrheology}
\end{equation}
where $\widetilde{\Delta \sigma^2}(\omega)$ and $\widetilde{\Delta\mathbf{r}^2}(\omega)$ represent the ensemble and time averaged Fourier power spectra of the stress and strain fluctuations respectively. To probe the strain fluctuations in the continuum, each bubble or particle is treated as a tracer moving in a viscoelastic continuum formed by all the other bubbles, which is driven by the fluctuating stresses acting upon it \cite{G*Mason2000,MicroRheoLubensky2003}. The random fluctuating stress acting on each bubble is computed with a virial method \cite{StressTesta2010, StressZhou2003}. 

\begin{figure}[t]
\centering
\begin{captivy}{\includegraphics[width = \linewidth]{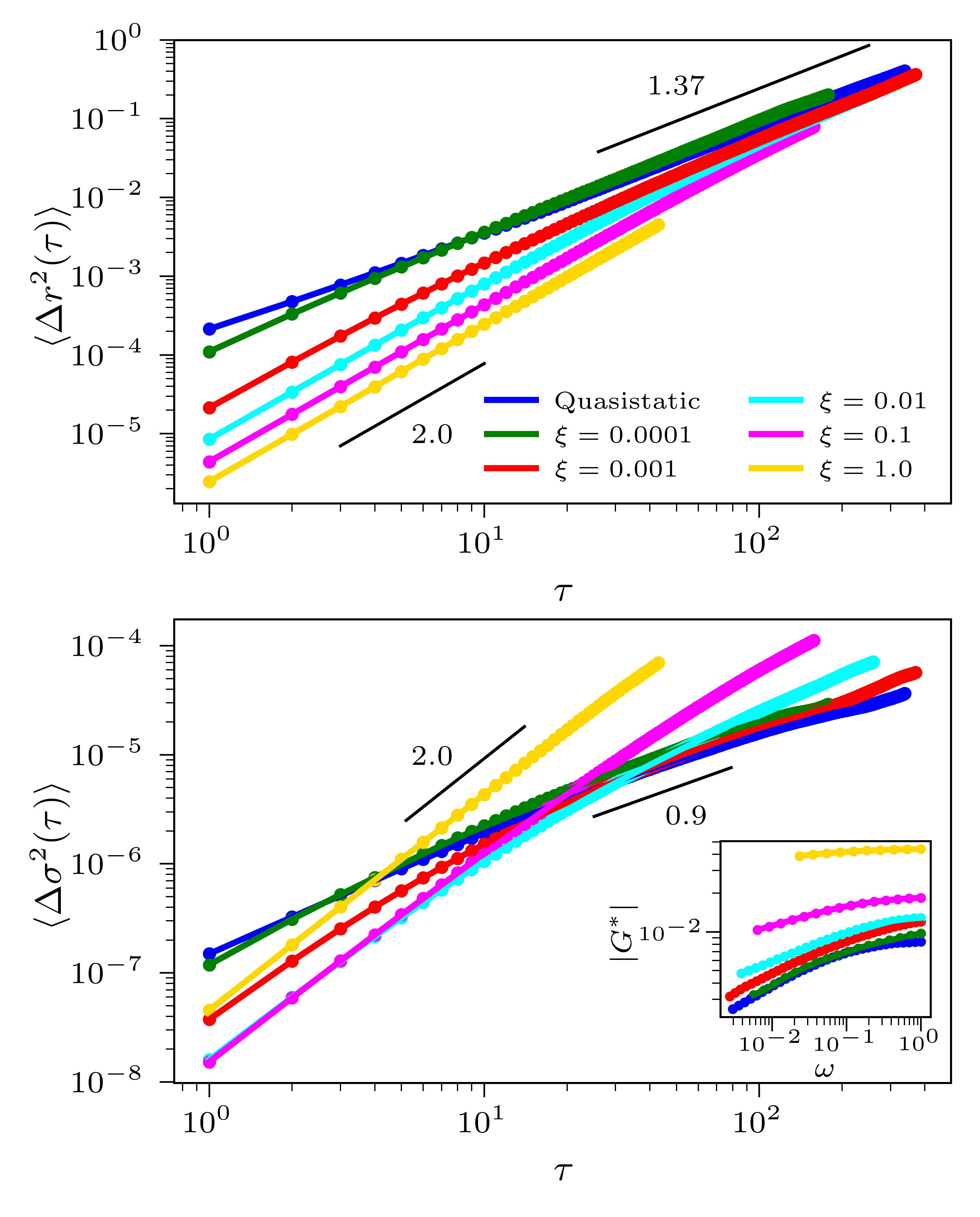}}
    \oversubcaption{0.05, 0.98}{}{fig:MSDsa}
    \oversubcaption{0.05, 0.48}{}{fig:MSDsb}
\end{captivy}
\caption{{\textbf{Mean-squared displacements and stress enable the characterization of rheology}}:
(\textbf{a})  Mean squared displacement plotted for the different $\xi$ values shows ballistic motion rolling over to a super-diffusive form for lower $\xi$ simulations. In comparison, more viscous simulations show more ballistic behavior over a larger range of $\tau$. The short lines are eye-guides to compare to the logarithmic slope of the curves.
(\textbf{b}) Mean squared stress plotted for the different $\xi$ values shows ballistic motion rolling over to a weakly sub-diffusive form for lower $\xi$ simulations. In comparison, more viscous simulations show more ballistic behavior over a larger range of $\tau$. We note here that the roll-over to sub-diffusive character appears sensitive to finite size effects \cite{SGMHwang2016} and poor statistics. Thus, we cut-off our data at long $\tau$ (using a statistical criteria, see \Cref{subsecSI:MSDfractal} - trimmed data shown above), to avoid such effects.
(\textbf{inset}) The computed dynamic shear moduli have a low frequency power-law frequency dependence for low and intermediate $\xi$ or viscosity. Higher viscosity systems cross over to more elastic behavior.}
\label{fig:MSDs}
\end{figure}

Given that the stress and strain signals for each bubble are random functions of time, it is more convenient to compute the ensemble and time-averaged squared amplitude of these quantities in the form of their lag time dependent mean squared differences, or MSDs. The conventional MSD (or strain MSD) is computed by pooling the positional motion in all three dimensions, while the stress MSD is computed by pooling the three off-diagonal (shear) elements of the bubble-wise symmetric virial stress tensor (see \hyperref[sec:Methods]{Methods} for details). The strain and stress MSDs of our simulated foams are shown in \Cref{fig:MSDs} for a range of damping values $\xi$. In the low viscosity and quasistatic limit, the strain MSD is super-diffusive over most of its lag time range, $\langle {\Delta \mathbf{r}^2(\tau)} \rangle \sim \tau^a$ with $a \approx 1.36$. For details of how the exponents are fit, see \Cref{subsubsecSI:slope}. As $\xi$ increases the short lag time behavior shows an increasingly ballistic or persistent form before crossing over to the previous super-diffusive behavior at long times. For the stress MSDs, the low viscosity and quasistatic case shows a diffusive to weakly sub-diffusive behavior, $\langle {\Delta \sigma^2(\tau)} \rangle \sim \tau^b$ with $b \approx 0.88$. As $\xi$ increases, the initial ballistic motion persists for longer times before appearing to cross over to the same sub-diffusive behavior at long times. Since the MSDs are computed by ensemble averaging over a very polydisperse sample, we also verified that the observed asymptotic power-law exponents are not significantly dependent on bubble radius, see \Cref{subsubsecSI:msdreadiusdep}.

To deduce the viscoelasticity of our different simulation cases from these MSDs, we used a method \cite{G*Mason2000} that approximates the power-spectra using the logarithmic derivative of the MSDs, see \Cref{subsubsecSI:calculatingFTs}.  This analysis yields dynamic shear moduli shown in the inset to \Cref{fig:MSDsb}. These curves display a power-law dependence at low frequencies $|G^\ast(\omega)| \sim \omega^{0.3}$ similar to that seen in real foams \cite{ViscoelasticVeronique2022, FoamsClary2022}. 
 The rheological response crosses over to elastic ($\beta = 0$) at the highest $\xi$ values, consistent with other experiments \cite{FoamGopal2003, RheologyEmulsionDutta2016}. 

When doing quantitative microrheology, a key factor is mechanical homogeneity—--the $3 \pi a$ factor in \Cref{eq:microrheology} assumes a spherical tracer with radius $a$ and no-slip boundaries in a homogeneous continuum. Often heterogeneity can cause deviations from this scaling, while still permitting the correct frequency dependence to be determined. Examining the radius dependence of our strain MSD amplitudes, see \Cref{subsubsecSI:hetergeneity} and \Cref{figSI:dr2a}, reveals slight $\textit{De}$-dependent deviations from $1/a$ scaling, especially for smaller bubbles and when viscous forces dominate. A similar analysis in one experimental system also showed near continuum scaling \cite{FoamsClary2022}, while in another, a weaker $a$-dependence was observed \cite{AnomalousGuidolin2024}.  Like both of those studies, we find similar lag time dependence of MSDs across bubble sizes, indicating that the computed shear modulus is expected to have the correct frequency dependence.

\begin{figure}[t]
\centering
\begin{captivy}{\includegraphics[width = \linewidth]{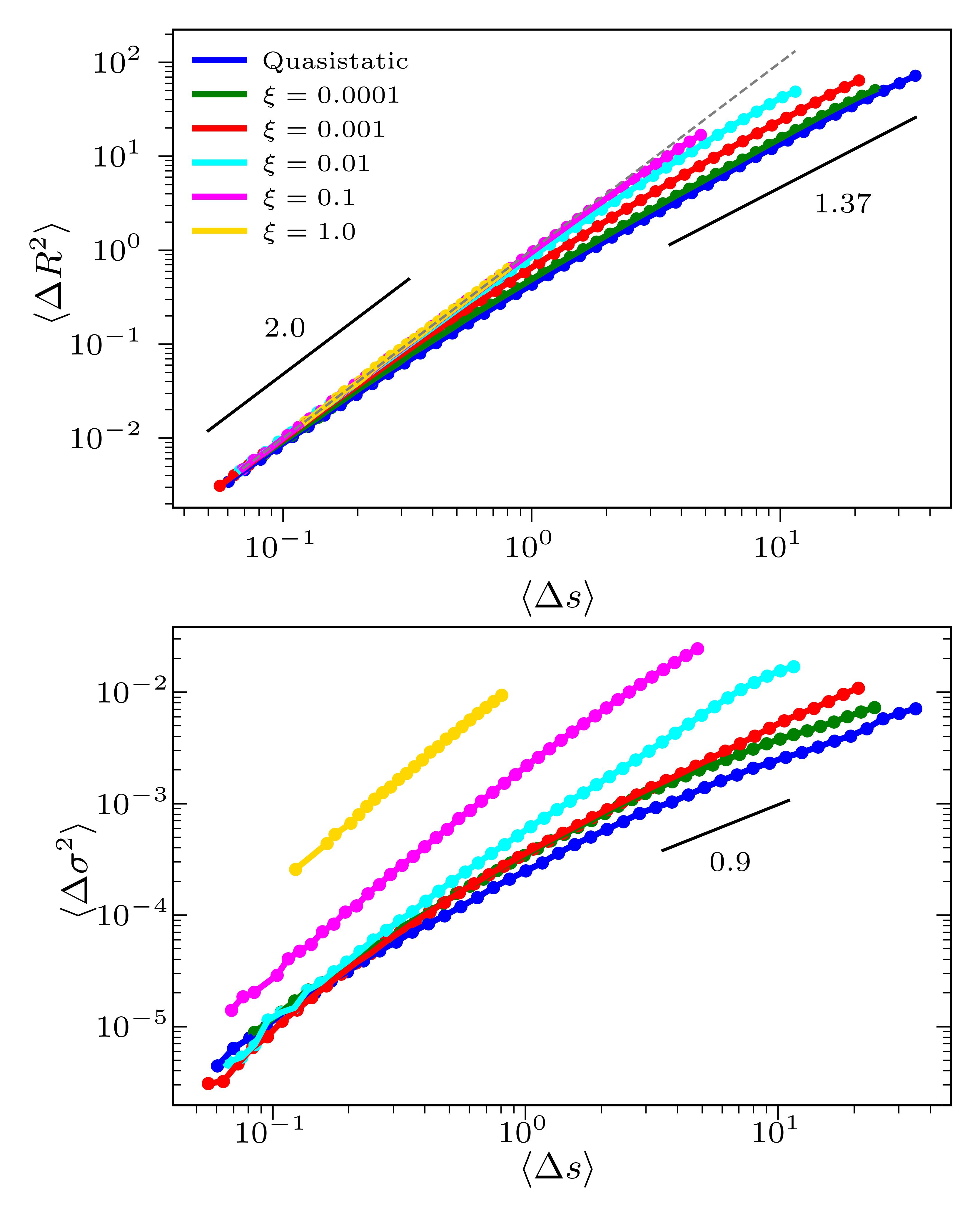}}
    \oversubcaption{0.05, 0.98}{}{fig:stressMSDsa}
    \oversubcaption{0.05, 0.48}{}{fig:stressMSDsb}
\end{captivy}
\caption{\textbf{Configuration space paths determine power-law rheology}: 
The high-dimensional mean-squared displacement (\textbf{a}) or strain MSD and mean-squared shear stress displacement (\textbf{b}) or stress MSD as a function of damping $\xi$ display forms that are very similar to the same measures shown in \Cref{fig:MSDs}. Unlike those measures, these are a function of contour length differences, $\langle \Delta s \rangle$, making them a reflection of the path's self-similar shape and energy function. This shows that the low frequency power-law rheology seen in the previous section is a consequence of fractal landscape geometry, and not a dynamical effect.} 
\label{fig:stressMSDs}
\end{figure}

\section{\label{sec:whyrheology} Power-law Rheology is due to a Fractal Landscape}

The question as to why foams have power-law rheology amounts to the question of why the strain and stress MSDs have the form observed above. In previous publications \cite{SGMHwang2016, FoamsClary2022}, we showed that the dynamics of quasistatic foams are closely tied to the geometry of the potential energy landscape of the foam spanning the $3N$-dimensional configuration space of $N$ bubbles. In particular, system configurations evolve via steepest descent on the landscape, and the corresponding paths were found to be self-similar (or fractal) with an emergent fractal dimension that could be algebraically related to the super-diffusive and rheology exponents $a$ and $\beta$, respectively. We will reprise and expand upon this analysis for the data in this study to explain the origin of power-law rheology in the finite $De$ case.

To characterize the high-dimensional path taken by the evolving system, we consider pairs of $3N$-dimensional configurations at different time points and compute their Euclidean distances $(\Delta R^2)$ and the contour length $(\Delta s)$ of the path connecting them (see \Cref{figSI:dr2ds}). These values can be pooled and ensemble averaged, yielding a function in $3N$-d conceptually similar to the conventional (strain) MSD in $3$-space, but with the lag time $\tau$ replaced with the contour distance $\Delta s$, see \hyperref[sec:Methods]{Methods} and \Cref{subsecSI:MSDfractal} for details. The results of this analysis are shown in \Cref{fig:stressMSDsa} for several values of $\xi$. Notably, the shape of the different $\xi$ curves in \Cref{fig:stressMSDsa} are very similar to the strain MSDs shown previously in \Cref{fig:MSDsa}. This similarity is not a coincidence; because $\Delta s$ and $\tau$ are highly correlated (see \Cref{figSI:dsdt}), the form of the strain MSDs versus lag time must have a very similar form to the high-dimensional curves.  

Unlike an MSD however, the form of these `high-dimensional MSDs' do not reflect dynamics (a function of $\tau$), but rather a geometrical measure (a function of length scale $\Delta s$) of the paths' tortuosity. Indeed, power-law scaling of such high-dimensional MSDs, $(\Delta R^2) \sim (\Delta s)^{1.37}$ indicates fractal scaling, with a fractal dimension given by $D_f = 2/1.37 = 1.46$, for the configuration path \cite{SGMHwang2016}. In this way, we have argued that the super-diffusive form of the strain MSD in the quasistatic case is simply a reflection of the self-similar geometry of the potential energy landscape \cite{SGMHwang2016, FoamsClary2022}. 

In this study, the addition of viscous stresses changes the picture slightly, because the configuration path no longer need follow the descent paths of the potential energy landscape. Notably, fractal scaling is still observed even at intermediate viscosities, where the dynamics are non-intermittent and the configuration is not obviously affected by energy minima. This finding is at first puzzling, if one assumes that the self-similarity of the path is due to the energy minima sampled during ripening and their distribution in configuration space. Rather, it appears that the fractal geometry of the path is due to the geometry of landscape itself, which will be investigated further in the following section.  Moreover, with increasing $\xi$ a ballistic regime appears for small $\tau$ and $\Delta s$ in the conventional and high-dimensional strain MSDs. The crossover from ballistic ($D_f=1$) to fractal scaling moves to progressively larger length-scales $\Delta s$ with increasing damping $\xi$, before being limited by the length of the simulations at higher $\xi$. Hints of the fractal rollover can still be observed, however, for $\xi$ values $\gtrsim 0.01$.  We interpret these findings as indicating that the influence of the fractal landscape is still felt on the longest length scales $\Delta s$ and timescales $\tau$, while viscous stresses dominate on shorter scales and produce persistent motion. But importantly, since $\tau$ and $\Delta s$ remain proportional, the shapes of the $3$-d and $3N$-d MSDs remain the same here as well.

A very similar situation appears to prevail for the stress MSDs. \Cref{fig:stressMSDsb} shows a high-dimensional analysis that corresponds to the stress MSD introduced in the previous section, but plotted versus the contour distance $\Delta s$ between pairs of configurations. Notably, for the quasistatic case, the curve scales as a power-law, $\langle \Delta \sigma^2 \rangle \sim (\Delta s)^{0.9}$, for large length scales (see \Cref{subsubsecSI:slope} for details). Here again, a power-law scaling of this function indicates that the magnitude of the forces and stresses associated with the energy landscape display self-similarity on the corresponding range of length-scales. As with the previous case, increasing $\xi$ leads to a short-range `ballistic' regime, which indicates that the stresses are persistent on those lengthscales. The shape or form of the curves in \Cref{fig:stressMSDsb} correspond, as before, to those in \Cref{fig:MSDsb}, demonstrating that as with the strain MSDs, the form of the stress MSD is a projection of the high-dimensional stress MSD, and in turn, controlled by the self-similar geometry of the landscapes gradients (forces and stresses) at low $\xi$ and long length scales.

Taken together, we can now see the origin of power-law rheology in the \textit{fractal landscape dynamics} picture \cite{SGMHwang2016,FoamsClary2022}. That is, the self-similar geometry of the energy landscape leads both the stress and strain MSDs to scale as power-laws at long $\tau$: $(\Delta r^2) \sim \tau^a$ and $(\Delta \sigma^2) \sim \tau^b$. Since the temporal power-spectra of the fluctuating stresses and strains are simply the Fourier transform of the respective MSDs due to the Wiener-Khinchin theorem, this implies that $(\widetilde{\Delta r^2}) \sim \omega^{-(a+1)}$ and $(\widetilde{\Delta \sigma^2}) \sim \omega^{-(b+1)}$ at low frequencies. Substituting these scalings into \Cref{eq:microrheology} then implies that $G^*(\omega) \sim \omega^\beta$ with $\beta = (a-b)/2$  and $\beta = 0.24 \pm 0.05$ for the exponents we observe. This value compares favorably to experimental values of $\beta = 0.17 \pm 0.03$ \cite{ViscoelasticVeronique2022} and $\beta = 0.19 \pm 0.03$ \cite{FoamsClary2022}. The same exponent argument in the high frequency limit yields an elastic plateau, since both MSDs scale as $\tau^2$ for small $\tau$; this high-frequency plateau can be seen in the full numerical analysis shown in \Cref{fig:MSDsb}(inset). This elasticity of landscape-driven foams is different from that of a jammed solid resting at a potential energy minimum, and is instead a property of a dynamic structure in mechanical equilibrium between viscous and interaction forces.

A key limitation of the rheology calculations above is the short duration of our simulations, limited by large bubbles ripening to the point of forming multiple contacts. It should be noted that the maximum value of the MSDs in \Cref{fig:MSDs}(a) indicates that the typical bubble moves only a fraction of the mean bubble radius $\langle a \rangle \approx 0.6$ during the course of the entire simulation. This motion limitation is even more severe at higher $\xi$. It thus remains possible that structural rearrangements or dynamical heterogeneity \cite{FoamsClary2022, MatrixGuidolin2024} 
at displacement scales beyond what we can observe could plausibly lead to different rheological behavior at longer times, such as a terminal relaxation mode.

\section{\label{sec:exploring} Exploring the energy landscape without ripening}

While the analyses in the previous sections are illuminating regarding the origin of foams' super-diffusion and power-law rheology, they also leave several questions unanswered.  For one, the $3N$-dimensional potential energy landscape is time-dependent due to ripening, could some of the observed landscape-driven dynamics be affected by its time-dependence? Similarly, how the system explores the landscape due to ripening is presumably biased, could the observed self-similarity only be a property of the landscape sampled in that biased fashion? Lastly, how do the sampled energies of viscous foams compare to the typical energies in thermalized systems in the super-cooled and glass regimes?

\begin{figure}[t]
\centering
\begin{captivy}{\includegraphics[width=\linewidth]{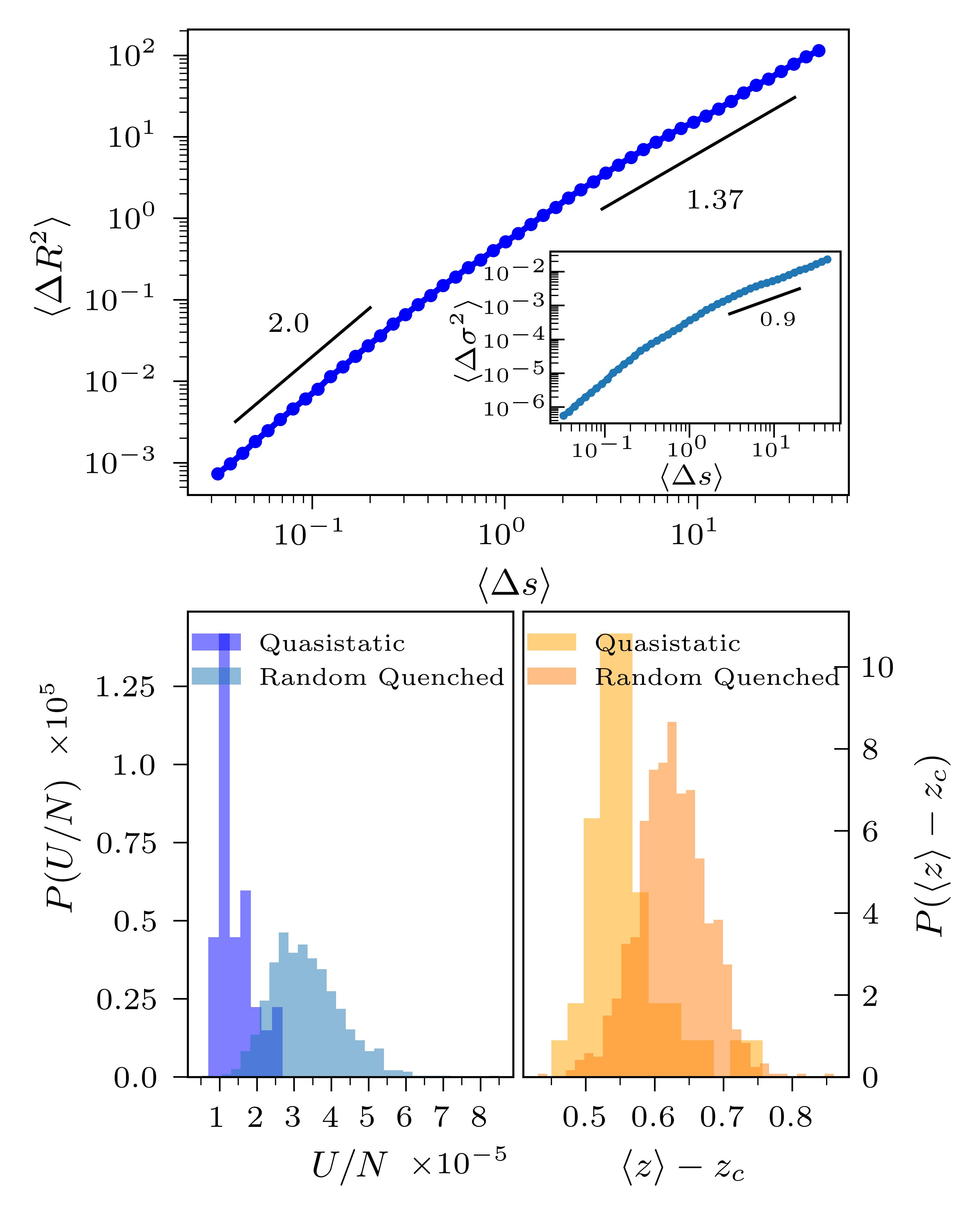}}
    \oversubcaption{0.05, 0.98}{}{fig:sqrheologya}
    \oversubcaption{0.05, 0.50}{}{fig:sqrheologyb}
    \oversubcaption{0.97, 0.50}{}{fig:sqrheologyc}
\end{captivy}
\caption{\textbf{The random quenched RQ and quasistatic ripening QR models are similar}:
(\textbf{a}) The high-dimensional Euclidean displacement versus contour distance, as in \Cref{fig:stressMSDsb}, for the RQ model exhibits fractal scaling very similar to the QR simulation. 
(\textbf{inset}) The stress scaling also shows similar fractal scaling to the QR case. (\textbf{b}) Comparing the average bubble energy $U/N$ for configurations in the RQ and QR ensembles shows that RQ minima have energies resembling that of quenched super-cooled fluids and glasses, as explained in the text. (\textbf{c})The mean coordination number $\langle z \rangle - z_c$ (where $z_c = 6$, is the critical coordination number) shows very slightly lower coordination of the QR ensemble than the RQ ensemble.}
\label{fig:sqrheology}
\end{figure}

To address the questions above, we study systems that are initialized at random points in configuration space and then evolved by FIRE energy minimization without ripening, which we term the `randomly quenched' (RQ) model.  To assess the landscape self-similarity, as before we compute the the high-dimensional stress and strain MSDs, shown in \Cref{fig:sqrheologya}.  Notably, the results resemble those seen in quasistatic ripening foams at lower energy. In particular, these curves show very similar asymptotic power-law scaling exponents ($\langle {\Delta \sigma^2(\tau)} \rangle \sim \tau^a$ with $a \approx 1.35$ and $\langle {\Delta \sigma^2(\tau)} \rangle \sim \tau^b$ with $b \approx 0.93$) as the quasistatic case at both long and short lengthscales, $\Delta s$. The strain MSDs show simple crossovers at similar lengthscales. One notable difference is that the magnitude of the stress MSD is a couple orders of magnitude larger than in the quasistatic case, and its crossover shifts to slightly longer lengthscales. This difference could be naturally explained by the magnitude of the potential gradients being proportionally higher at the higher potential energies being explored. 

The similarity of the two models' results leads to three immediate conclusions. First, the observed self-similarity that gives rise to FLD appears to be a generic feature of much of the energy landscape, in particular in the range of energies explored by the RQ model (which overlaps the range of the viscous foam model, and extends to slightly higher energy). Second, since the ballistic MSD behavior seen with increasing \textit{De}, as in \Cref{fig:stressMSDs} (at progressively higher energies in the landscape) is not seen on the RQ model it must be due to viscous stresses or biased sampling of the landscape due to elevated viscous stress. We hypothesize that the ballistic motion is due to the larger viscous stresses at high \textit{De} dominating the effects of small-scale undulations of the energy landscape. Third, the similarity of ripening and non-ripening models leads us to conclude that the large scale structure of the landscape must remain effectively unchanged due to ripening over the duration of our simulation.

Beyond the conclusions above, the similarity between the three models---quasistatic (ripening), viscous (ripening) and randomly quenched (no ripening)---remains puzzling. How can the long-range geometry of the configuration path resulting from different driving and equations of motion be effectively identical?  We hypothesize that the similarity is the consequence of all three models following a \textit{biased} energy minimizing path over the same energy landscape. Ripening induces a strain field on the system pushing the configuration in an essentially random direction in configuration space. Slowly changing bubble radii cause shallow energy minima to become unstable at a finite rate \cite{SGMHwang2016, Sessoms2010unexpected}, allowing the quasistatic system to evolve along an energy minimizing pathway, while also biased to evolve in the direction preferred by the ripening strain. Similarly, a viscous foam with ripening will evolve balancing viscous and potential forces, \Cref{eq:damped-equation}, while also biased to move in the direction dictated by the ripening strain. We suppose that this motion is dominated by the same self-similar landscape structures as the quasistatic case at long length-scales, while the configuration `bypasses' energy minima to evolve continuously. The randomly quenched model, without ripening, feels a strong potential energy gradient (due to a few large bubble overlaps) that also picks a random direction in configuration space. The RQ system then evolves according to an energy minimizing pathway biased by that global potential gradient direction, with its path being perturbed by large-scale landscape structures that resembles those seen in the ripening foams.

Lastly, we seek to compare the energy scales explored by quasistatic and viscous ripening foams to the energies associated with different super-cooled and glassy materials. To characterize the latter, we will follow the work of Sastry, Debenedetti, and Stillinger \cite{SimAnnealSastry1998}, who considered the distribution of energy minima (also called inherent structures) resulting from rapidly quenching thermalized states of glass forming liquids. We construct two ensembles with matched energies, one quasistatic ripening (QR) and the other random quenched (RQ), by matching the mean bubble size $\langle a \rangle$ and bubble number $N$. The quasistatic ripening, QR, ensemble of $\sim 50$ configurations consists of ripening simulations that occupy similar average radii (${\langle a \rangle}^2 \approx 0.49$; corresponding to the data in \Cref{fig:avalanchesc}) and a narrow range of $N$. The random quenched (RQ) ensemble comprises $1000$ configurations: for each QR configuration, we apply $20$ independent scrambles followed by FIRE minimization to locate local minima. The RQ ensemble in principle provides a random sampling of all the minima in the entire landscape, weighted by the catchment hypervolume\footnote[2]{A recent study \cite{MiragesSuryadevara2024} has indicated that earlier findings of self-similar catchment hypervolume distributions using FIRE relaxation are numerically suspect. This finding does not affect our essentially qualitative conclusions from the RQ ensemble. Other measures of self-similarity used in our work (e.g. MSDs of FIRE trajectories) were not implicated by this study.} of their associated energy basins \cite{BasinsXu2011,MiragesSuryadevara2024}.

Comparing the two ensembles in terms of mean energy per particle and mean coordination number (see \Crefsubfiglist{fig:sqrheology}{b}{c}) reveals broad, overlapping distributions. The mean energy of the RQ states is considered to correspond to that of a fluid quenched from above the onset temperature \cite{SimAnnealSastry1998}, where glassy phenomena first begin to manifest. Notably, the mean energy of the QR ensemble lies near the lower bound of the RQ distribution. Since random quenching does not access the deepest glassy states, this overlap suggests that the QR ensemble spans a broad range of configurations—from supercooled fluid states up to the onset temperature to configurations consistent with moderate thermal annealing. This range may help explain why the dynamical heterogeneity seen in slowly ripening foam models \cite{FoamsClary2022,MatrixGuidolin2024} closely resembles that of supercooled colloidal suspensions \cite{ColGlassWeeks2000, DynamicalZhang2016}.

Finally, the RQ ensemble was constructed to correspond to the data for viscous ripening foams shown in \Cref{fig:avalanchesc}. This allows us to also contextualize our viscous foams using the Sastry, Debenedetti, and Stillinger \cite{SimAnnealSastry1998} framework. In this language, the energies of our intermittent foams roughly map onto their `landscape-dominated' regime, our landscape-driven foams with power-law rheology correspond to their `landscape-influenced' regime, and our viscous-driven foams with elastic response resemble their high-temperature, pre-onset regime.  It should be noted, however, that these comparisons are of the foams' energies to the inherent structure energy of thermalized systems, not their mean energy.

\begin{figure}[t]
\centering
\begin{captivy}{\includegraphics[width=\linewidth]{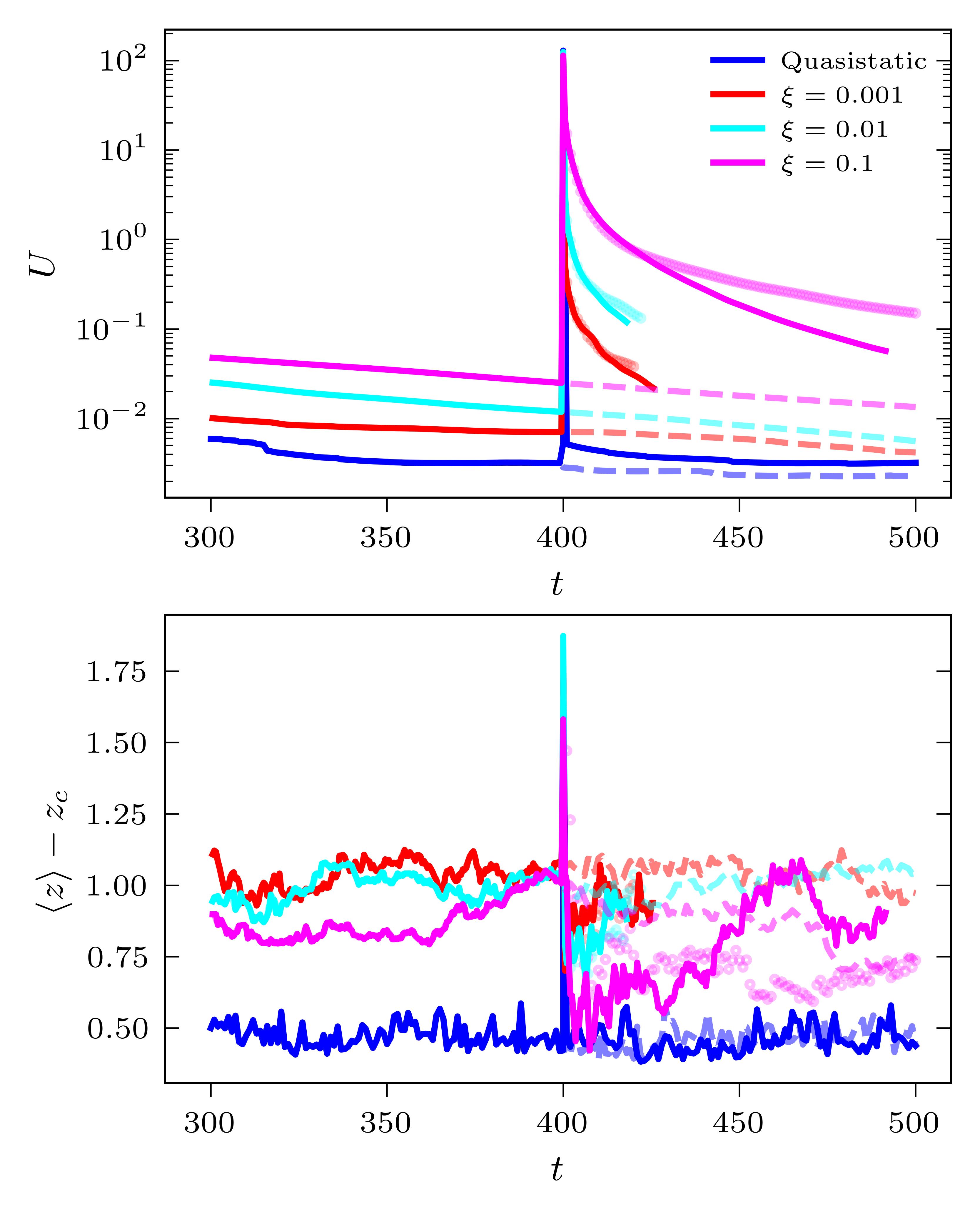}}
    \oversubcaption{0.05, 0.98}{}{fig:reseta}
    \oversubcaption{0.05, 0.48}{}{fig:resetb}
\end{captivy}
\caption{\textbf{Scrambled foams with damping show a very slow relaxation towards their prior trends in energy and coordination.} Finite $\xi$ simulations are scrambled at $t=400$ and evolved using \Cref{eq:damped-equation} (light colored curves). Dashed lines depict the same simulation without scrambling as a baseline. Data shown above are from the same initial configuration except at $\xi = 0$ where a different configuration is used (see \Cref{subsecSI:a2t} for details).
(\textbf{a}) The potential energies post scrambling require progressively more time to recover their steady state. This relaxation process proceeds slightly faster in the absence of ripening (darker colored curves), confirming that relaxation is not ripening dependent.
(\textbf{b}) The mean coordination number $\langle z \rangle$ of the system shows similar $\xi$ dependent trends. Interestingly, the initial dynamics directs the system to lower $\langle z \rangle$ configurations before relaxing to the appropriate steady-state value.}
\label{fig:reset}
\end{figure}

\section{\label{sec:recovery}Slow recovery after strain perturbation due to viscous relaxation}

Experiments have observed \cite{PerturbfoamsGopal1995, MemoryFoamHohler1999} that foams perturbed by large-strain motions relax back to their unperturbed steady state, but only after an unexpectedly long waiting time, $\tau_W$. The time for relaxation to the nearest state of mechanical equilibrium was typically assumed to be the viscous times scale, $\tau_{\xi} = \xi {\langle a \rangle}^2/ \epsilon$, which is orders of magnitude shorter than the observed waiting time for recovery \cite{FoamnostrainBuzza1996, FoamnostrainBouchama1996}. These observations were broadly taken as evidence for glassy relaxation in foams; the idea being that mechanical perturbation places the foam in an atypical state and recovery requires ripening-induced activation over numerous energy barriers \cite{PerturbfoamsGopal1995, MemoryFoamHohler1999}.  Given that foam mechanics appear to be due to fractal landscape dynamics and not hopping amongst glassy states, the true origin of the slow relaxation remains to be elucidated. 

We performed simulations using our damped ripening foam model for various $\xi$ values subjected to mechanical perturbation. To simulate mechanical perturbation such as stirring or large shear deformation, the positions of the bubbles were randomized at a selected point in time. The damped systems are then evolved according to the relaxation-ripening procedure described previously in \Cref{sec:bubblemodel}. For the quasistatic case ($\xi = 0$), the perturbed system is relaxed to its first energy minimum using FIRE \cite{FIREGumbsch2006} instead of using \Cref{eq:damped-equation}. 

\begin{figure}[t!]
\centering
\begin{captivy}{\includegraphics[width=\linewidth]{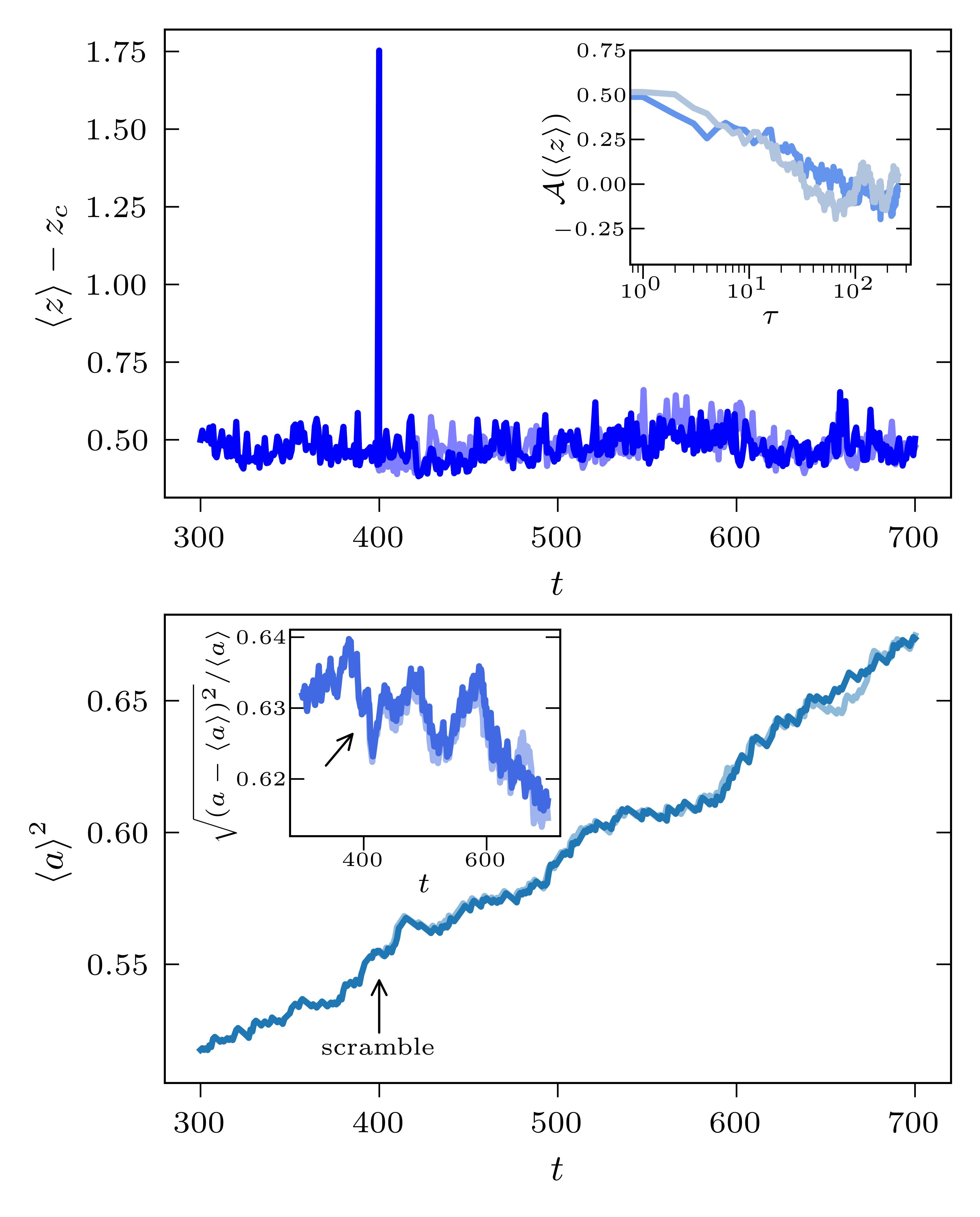}}
    \oversubcaption{0.05, 0.98}{}{fig:QSreseta}
    \oversubcaption{0.05, 0.48}{}{fig:QSresetb}
\end{captivy}
\caption{\textbf{Scrambling a quasistatic foam does not alter multiple measures of ripening dynamics.}
(\textbf{a}) We scramble the configurational positions of quasistatic foam at $t = 400$), dark curve, and compare the result to the unscrambled trajectory (light curve). Neither the mean coordination versus time nor its temporal autocorrelation (\textbf{top inset}) show a significant perturbation.  (\textbf{b}) Similarly, the squared mean bubble radius does not significantly change its evolution (linear as expected for dynamical scaling) upon scrambling, nor does the standard deviation (\textbf{bottom inset}) of the bubble radius distribution. All data in this figure was generated using a single initial configuration.
}
\label{fig:QSreset}
\end{figure}

The results of our perturbation experiments on ripening foams at different damping parameters are shown in \Cref{fig:reset}. As seen in the experiments \cite{FoamGopal2003, MemoryFoamHohler1999}, the systems' energies shown in \Cref{fig:reseta} require a surprisingly long time to recover to their former steady state trend and at high damping $\xi$ do not have time to recover before ripening ends the simulation. There is a qualitative correlation between the simulated viscosity and the rate of energy relaxation. \Cref{fig:resetb} shows the corresponding progression of the average coordination number toward steady state; the $z$ values transiently drop below the steady-state line after perturbation, which could be consistent with the change of rearrangement rate seen in experiment \cite{PerturbfoamsGopal1995}. The relaxation of perturbed foams corresponds to the system evolving on high energy, often minima-free portions of the landscape at a speed which is limited by the viscous damping parameter. The lack of energy barriers is confirmed by the quasistatic curves in \Cref{fig:reset} returning to the steady state after a single FIRE relaxation. Even more striking, repeating the relaxation simulations again with ripening `turned off' after the scramble step, as shown in \Cref{fig:reset}, shows that recovery actually appears slightly faster, confirming that ripening is not required for relaxation at all. 

Landscape geometry provides an explanation for the waiting time for recovery being much longer than the time for viscous relaxation: $\tau_W / \tau_{\xi} \simeq \mathcal{O}(10^3)$. We consider that $\tau_{\xi}$ is the time for the system configuration to traverse between two neighboring energy minima at a viscosity limited rate, while $\tau_W$ is the time to traverse between a random point in configuration space to the corresponding landscape basin minimum, also at a viscosity limited rate. Our previous work has shown that energy minima are not randomly distributed in configuration space, but rather formed into dense and fractal clusters \cite{SGMHwang2016, FoamsClary2022}. As a result the high-dimensional distance between two neighboring minima (within a cluster) is far smaller than the distance from a random configuration to even the closest energy minimum. Indeed, FIRE relaxations (or steepest descent) from random points in configuration space typically require $\mathcal{O}(10^5)$ steps and a correspondingly large displacement in configurations space. Conversely, FIRE relaxation after a ripening move requires far fewer steps, and traverses a much smaller distance. The slow recovery in foams is thus a consequence of the clustering of minima in the energy landscape.

Remaining open questions regarding the mechanical perturbation of foams include \textit{i}) to what extent the perturbations alter the structure and dynamics of the quasistatic ripening system, if at all, and \textit{ii}) whether perturbation alters the ripening process, i.e.~the time-dependent bubble size distribution in dynamical scaling.  While one prior study \cite{zcoarseningHilgenfeldt2001} found a change in ripening dynamics due to changes in coordination number and structure, we do not observe such effects. \Cref{fig:QSreseta} shows that the mean coordination of the quasistatic ripening system does not change upon positional scrambles. To screen for possible changes to the dynamics, we compute the temporal autocorrelation function of the fluctuating mean coordination and find that the systems with and without positional randomization show very similar, roughly logarithmic decay profiles. Moreover, analysis of the time-dependent bubble size distribution also shows no significant effects as a result of randomization. Measures, including the first and second moment of the radius distribution, show no significant change from steady-state behavior, see \Cref{fig:QSresetb}, consistent with earlier experimental reports \cite{PerturbfoamsGopal1995}. This finding seems to eliminate the possibility that slow relaxation in foams is somehow related to the potentially slow dynamics of the bubble size distribution or the approach to dynamical scaling.

\section{Synthesis and Outlook}

\begin{figure}[t]
\centering
\begin{captivy}{\includegraphics[width=\linewidth]{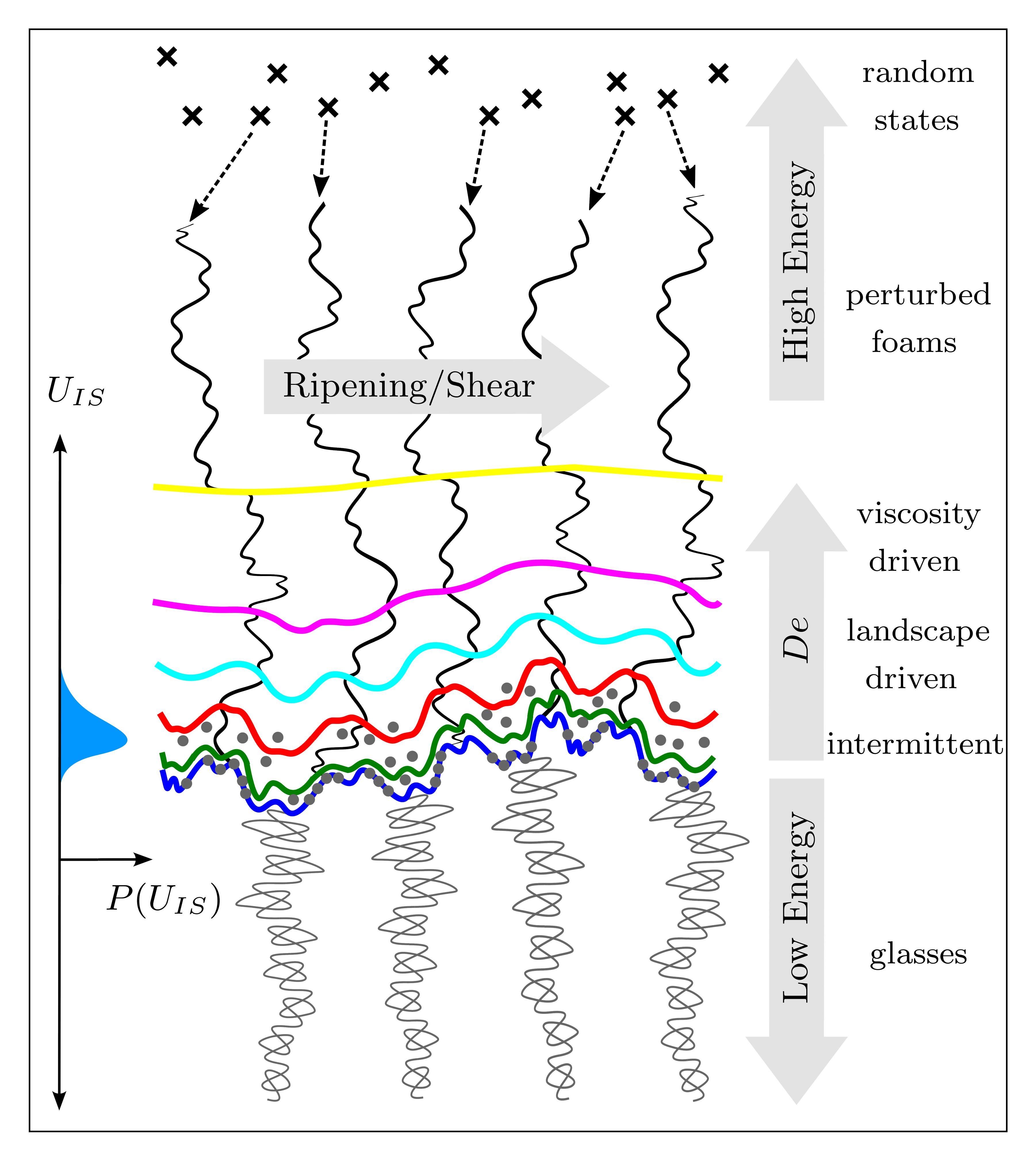}}
\end{captivy}
\caption{\textbf{Schematic of different configuration space paths as described in the text}: An 'aerial' view of different observed paths on the energy landscape, higher energies at the top, lower at the bottom.  Ripening (or shear) propels the configuration across the landscape (left to right) at a steady state energy that depends on Deborah number, \textit{De} (horizontal paths in center, colored as elsewhere). Random configurations (crosses) relaxed by FIRE descend in energy (solid black vertical curves) forming curves with the same fractal geometry, $D_f \approx 1.45$, as the quasistatic ripening path (blue). These RQ paths terminate at inherent structures (gray circles) with potential energy $U_{IS}$, having a probability distribution $P(U_{IS})$, left.  A recent publication \cite{MIMSEThirumalaiswamy2022} shows that very low energy configurations deep in the glass (not sampled by quenching) are arranged into low dimensional fractal clusters (gray lines) having a distinctly higher fractal dimension, $D_f \approx 2.4$. 
}
\label{fig:landscapescheme}
\end{figure}

In this study, we have examined the athermal, viscous dynamics of polydisperse soft-sphere foams as they explore their energy landscape. The varied configuration space paths from our different simulations are summarized schematically in \Cref{fig:landscapescheme}, which is organized with higher energy at the top, and lower energy at the bottom. Strain associated with ripening drives the system in a particular direction (horizontal in the schematic); applied shear presumably would have a similar effect. The foam evolves along a path aligned with that direction during ripening, balancing potential gradients and viscous forces according to \Cref{eq:damped-equation}.  At higher \textit{De} (higher viscosity, ripening rate or higher shear rate), the system follows a path at a higher steady-state energy, where larger potential gradients can balance the higher viscous stresses. These paths exhibit a fractal geometry on large length-scales, which is due to the geometry of the underlying energy landscape. As \textit{De} increases, the path becomes more straight (or persistent) due to increasing viscous stress dominating potential gradients. This persistent motion is more prominent at smaller length-scales, and the cross-over to landscape-driven motion moves to longer lengthscales and lag times as \textit{De} increases.

Dynamics and rheology are also determined by the landscape geometry, which remains effectively static on long length-scales despite ripening. Low \textit{De} foams move intermittently from near one energy minimum associated with one jammed configuration to another.  These foams are not permanently arrested because energy minima are destabilized at a steady rate by ripening \cite{SGMHwang2016}, rather than a thermal-like activation process as supposed by SGR theory. As \textit{De} increases, the dynamics becomes continuous because the path no longer encounters energy minima that transiently arrest the dynamics, even though such minima are present at the energies these landscape-driven foams explore. In both cases, the super-diffusive dynamics result from a projection of their fractal configuration space path to real space.  Both cases also display power-law rheology as required by \Cref{eq:microrheology}, because the power spectrum of the fluctuating stresses is also self-similar. The straight line motion on short length-scales and higher \textit{De} lead to elastic rheological responses, explaining the varied rheology measurements in the literature.

The energy scales explored by foams can be compared to that of supercooled fluids and glasses using the distribution of inherent structure energies, $P(U_{IS})$, corresponding to the energy minima resulting from quenching fluids above the onset temperature, shown schematically in \Cref{fig:landscapescheme}. Low \textit{De} foams occupy a broad range of energies below the peak of $P(U_{IS})$, which correspond to the inherent structures resulting from quenching super-cooled fluids and moderately annealed glasses \cite{SimAnnealSastry1998}. Thus, while the microscopic dynamics of these materials does not resemble glass-like activated hopping over a distribution of energy barriers, intermittent foams certainly do display structural signatures of glassiness as well as some dynamical features such as dynamical heterogeneity.  Lastly, the slow relaxation of mechanically perturbed foams is not due to their being trapped in energy minima---as such minima are effectively absent at high energies. Slow relaxation is due instead to the long path they must follow (at a viscosity limited rate) to return to their previous steady state.

Additional studies show that the above picture of the energy landscape is not complete, and provide clues regarding why slowly deformed foams do not explore the lowest possible energies.  Polydisperse soft-sphere systems above $\phi_J$ have been shown to also possess a rich set of glassy and jammed states at much lower energies than $P(U_{IS})$ when relaxed using advanced sampling methods such as swap Monte Carlo \cite{swapHSBerthier2016} or modified athermal metadyanamics \cite{MIMSEThirumalaiswamy2022}. Our study \cite{MIMSEThirumalaiswamy2022} found these minima had very small catchment hypervolumes and lie along paths with a distinctly higher fractal dimension ($D_f \gtrsim 2.4$) than those higher in the landscape ($D_f \approx 1.45$), shown schematically at the bottom of \Cref{fig:landscapescheme}. This suggests that the landscape is multifractal---locally self-similar but having geometry that varies across different regions of configuration space.  We hypothesize that slowly deformed soft sphere systems do not access the deeper regions of the energy landscape because the relatively directed and straight paths they follow (due to ripening or shear strain) are geometrically incompatible with the far more tortuous and perhaps less branched energetic pathways at low energies. 

The physical situation described above seems closely analogous to that in deep learning models, where different optimizers are used to converge to different parts of the loss landscape, in particular regions that are flatter, smoother and thus more expressive \cite{DeepLearningKeskar2017}. The deep analogies between machine learning and glasses were established decades ago \cite{NNetGardner1988, SpinGlassMezard1987}, yet modern deep learning systems do not necessarily seek the deepest, glass-like minima of the loss landscape.  Recent studies \cite{AnomalousDiffChen2022, OptimizationLy2025} show that common deep learning optimizers tend to follow low fractal dimension paths that resemble the relaxation paths of foams when converging to more generalizable solutions and expressive regions of a multifractal loss landscape. In this viewpoint, while it could be said that ripening or sheared soft sphere systems may structurally and energetically resemble glassy states (at low \textit{De}), \textit{their dynamics are more analogous to training dynamics in expressive deep learning models}. Indeed, the ability of similar systems to develop trained mechanical responses using large amplitude oscillatory shear (LAOS) \cite{MemoryKeim2011, MechAnnealKeim2021, MechTuningEdera2025} is a manifestation of their steady state dynamics over an expressive (not glassy) potential energy landscape.

Future work will seek to understand the energy-dependent multifractal geometry of the soft sphere energy landscape using various forms of constrained and advanced energy minimization. How fractal landscape dynamics on the flatter, smoother higher-energy portion of the landscape can be mapped onto simple stochastic models such as Fractional Brownian Motion \cite{FBMMandelbrot1968} is also being explored. These models might provide a useful description of cytoskeletal dynamics \cite{LevySivarajan2024}, which have also been hypothesized to be related to learning dynamics \cite{NetworksGalvani2024, LearningBanerjee2025}. In the longer term, we expect that the analogy with deep learning models will also be fruitfully explored, and may provide insights for why both energy and loss landscapes have certain seemingly ubiquitous geometrical features.

\section{Methods}
\label{sec:Methods}
We focus our modeling efforts on the now canonical `bubble model' \cite{FoamMechDurian1995, BubbleDurian1997}. Here, the constituent bubbles are treated as independent soft-spheres that interact via a repulsive potential when they overlap, defined as follows.

\begin{equation}
  V(\mathbf{r}_{ij})=\begin{cases}
    \frac{\epsilon}{2} {\left(1 - \frac{\lVert{\mathbf{r}_{ij}}\rVert}{a_i + a_j} \right)}^{2}, & \text{if $\lVert \mathbf{r}_{ij} \rVert<a_i + a_j$}\\
    0, & \text{otherwise}.
  \end{cases}
\label{eq:soft-sphere potential}
\end{equation}

\noindent with $r_{ij}$ being the distance between two bubbles of radii $a_i$ and $a_j$, and $V(\mathbf{r}_{ij})$ being the corresponding potential.

We model the ripening mass flux between bubbles via \Cref{eq:mass-flux}, using the ripening parameters $\alpha = 0.05$ and $b_{mf} = 0.04$, chosen to correspond to our previous study \cite{SGMHwang2016}. When during system evolution a bubble's volume turns slightly negative, we remove it from the simulation box, while ensuring that the mass of the deleted bubble and its neighbors is conserved.

To integrate the equations dictating our system, we use a simple explicit Euler scheme with small $dt$ values, to numerically integrate the equations of motion (more details in \Cref{subsecSI:integration}. The use of other integrators, like a second-order Runge-Kutta numerical discretization, gave similar results. While we note that the equations of motion can be physically unstable at very high energies (when there is a significant overlap between bubbles), we verify that such large overlaps are not present at the energy levels simulated here. Furthermore, while using small step sizes within the range of numerical stability (more details in \Cref{subsecSI:integration}), we ensure the simulation has converged by cross-validating with smaller step sizes ($dt$). It may be noted here that smaller step sizes ($dt$) are required for lower values of $\xi$ and, therefore, are computationally more expensive to simulate.

Here, it must be noted for ensemble based calculations like the time and stress MSD, we choose a reduced ensemble of bubbles that are non-rattler having $4$ or more neighbors and remain 'alive' throughout the time span of the simulation, for our analysis in \Cref{fig:MSDs} and \Cref{fig:stressMSDs}. This keeps the number of dimensions ($3N_{red}$) used to calculate $\Delta R^2$ a constant, where $N_{red}$ is the reduced number of bubbles. The choice of the ensemble changes the terminal slope slightly, while preserving the overall qualitative trend.

We calculate the rheology of our system by analyzing the temporal strain and stress MSDs of individual bubbles. The local fluctuating stress on individual bubbles is defined as follows \cite{StressTesta2010, StressZhou2003}:

\begin{equation}
    \sigma(\mathbf{r}_i) = - \frac{1}{2} \left( \sum_{j}^{nn} \mathbf{r}_{ij} \otimes \mathbf{F}_{ij}\right) \delta (\mathbf{r} - \mathbf{r}_{i})
\label{eq:stress-equation}
\end{equation}

\noindent where $\mathbf{r}_{ij}$ and $\mathbf{F}_{ij}$ represent the distances and forces between particles $i$ and $j$, with the summation going over nearest neighbors. The MSDs obtained from the strain and stress are analyzed to obtain rheological moduli and exponents (for more details see \Cref{subsecSI:rheology})

\section*{Author Contributions}
Author contributions: A.T., R.A.R., and J.C.C. designed research; A.T. performed research and analyzed data; C.R-C. provided data analysis tools; A.T., R.A.R., and J.C.C. wrote the paper.

\begin{acknowledgments}
We are grateful for useful conversations with Andrea Liu and Talid Sinno. This work was supported by NSF-DMR 1609525 and 1720530.
\end{acknowledgments}

\appendix

\renewcommand{\thefigure}{A\arabic{figure}}
\setcounter{figure}{0}  

\section{\label{subsecSI:dampedSGM}Damped SGM Model}
\subsection{\label{subsecSI:underdamped}Underdamped limit of an overdamped equation}

The equation used for the simulations, as described in the main text, is:

\begin{align}
\begin{split}
    \xi \frac{d\mathbf{r}_i}{dt} & = \mathbf{F}_i \\
    & = -\mathlarger{\mathlarger{\sum}}_{j}^{nn} \frac{\partial V(r_{ij})}{\partial r_i}
\label{eq:damped-equation-2}
\end{split}
\end{align}

It may seen that the equation is similar to an overdamped equation of motion. However, while this equation resembles and has the characteristics of an overdamped equation of motion, it is not so due to large viscosity. This is rather due to the non-inertial nature of the constituent particles considered in the system. One may recall that dynamics for a mass attached to a damped spring are mediated by the damping factor $\zeta = b / (2\sqrt{km})$. That can be evaluated for our system of interest as follows $\zeta \simeq \xi /\sqrt{\epsilon \rho \langle a \rangle}$. Since overdamped dynamics is achieved when $\zeta \ge 1$, we see that the non-inertial particles $(\rho \to 0)$ in our case give rise to the so-called overdamped equation of motion above. Meanwhile, we continue to operate with a finite value of $\xi$.

\subsection{\label{subsecSI:integration}Integration and stability}

The simulation can be summarized as a numerical integration of the two equations -- \Cref{eq:damped-equation} and \Cref{eq:mass-flux} using a numerical integration technique. Due to the stiff nature of \Cref{eq:damped-equation} (especially at small $\xi$ values), one needs to choose appropriate $dt$ values to ensure any error perturbations don't diverge as the simulation proceeds, and  a converged solution is obtained. We use a simple Explicit Euler scheme to perform our integration here. We note that other methods, like implicit Euler and second-order Runge Kutta scheme, provide more extensive stability regimes for $dt$ and are more accurate but can have more significant computational overload associated with the integration scheme. Below, we perform a simple numerical stability test.

We start by considering \Cref{eq:damped-equation} for all $N$ particles or $3N$ degrees of freedom, i.e., $i \in \{1,2,...3N\}$, which can be expressed in terms of the Hessian by using a Taylor expansion as follows: 

\begin{align}
\begin{split}
    \xi \frac{d\mathbf{r}}{dt} & = \mathbf{F} \\
    & = \mathbf{F}_{0} - \mathbf{H} \mathbf{r}
\end{split}
\label{eq:damped-eq-approx}
\end{align}

where $\mathbf{r}$, $\mathbf{F}$ are $3N$ dimensional vectors and $\mathbf{H}$ is a $3N\times3N$ matrix or the Hessian of the potential field. We may note here that for low $\xi$ simulations, the system configurations are close to mechanical equilibrium, so for our stability analysis, we may approximate this using $\mathbf{F}_0 \simeq 0$. Further one may note that the any error $\epsilon_i$ would propagate via an equation similar to \Cref{eq:damped-eq-approx}:

\begin{align}
\begin{split}
    \xi \frac{d\mathbf{\epsilon}}{dt} & = - \mathbf{H} \mathbf{\epsilon}
\end{split}
\label{eq:damped-eq-epsilon}
\end{align}

Now, using the Explicit Euler formalism, for time steps $n+1$ and $n$, we get:

\begin{align}
\begin{split}
    \xi \frac{\mathbf{\epsilon_{n+1}} - \mathbf{\epsilon_{n}}}{dt} & = - \mathbf{H} \mathbf{\epsilon}_n \\
    & = -\mathbf{\lambda} \mathbf{\epsilon}_n \\
    \frac{|| \mathbf{\epsilon_{n+1}} ||}{|| \mathbf{\epsilon_{n}} ||} & = || \mathbf{I} - \mathbf{\lambda} || dt/\xi
\end{split}
\label{eq:damped-eq-epsilon-2}
\end{align}

where $\mathbf{\lambda}$ is a matrix containing all eigenvalues of $\mathbf{H}$. Enforcing the criteria of stability on the equation above we have,

\begin{align}
\begin{split}
    \frac{|| \mathbf{\epsilon_{n+1}} ||}{|| \mathbf{\epsilon_{n}} ||} & \leq 1 \\
    || \mathbf{I} - \mathbf{\lambda} || dt/ \xi & \leq 1 \\
    0 \leq || \lambda_{max} ||dt/\xi & \leq 2
\end{split}
\label{eq:damped-eq-epsilon-3}
\end{align}

where $\lambda_{max}$ is the largest eigenvalue for $\mathbf{H}$. Since all eigenvalues would be real for this physical system, we now have,

\begin{align}
\begin{split}
    0 \leq || \lambda_{max} ||dt/\xi & \leq 2 \\
    dt & \leq 2 \xi / \lambda_{max}
\end{split}
\label{eq:damped-eq-epsilon-4}
\end{align}

For most configurations, explored in our system simulation the $\lambda_{max}$ varies around $\sim 1-10$. This gives us that $dt \leqslant \xi/5$ as the condition for stability. Here, we choose $dt = \xi/ 10$, as the step size for all simulations reported in this study. Since we have \Cref{eq:mass-flux}, which also controls overall dynamics, this choice of time-step was validated for convergence. We have verified that our explicit Euler scheme was converged by checking other smaller values of $dt$. Other schemes like the RK-2 also produced similar results. While the above derivation only holds true for low $\xi$, we operationally verified that a similar criteria works for $\xi \sim 0.001-0.01$. Further, for $\xi > 0.01$, we stuck with the use of $dt = 0.001$, as the system moves further away from mechanically stable states and the above approximation in \Cref{eq:damped-eq-epsilon} fails to strictly hold.

\subsection{\label{subsecSI:Deborahnumber}Dimensionless group analysis: \textit{Deborah number}}

We evaluate the \textit{Deborah number} $De$ as the ratio of the damped relaxation time from \Cref{eq:damped-equation} ($\tau_{R} = \xi {\langle a \rangle}^2/ \epsilon$) and probing time associated with changing bubble radii $(a)$ imparted by the coarsening process, \Cref{eq:mass-flux} ($\tau_{C} = {\langle a \rangle}^2/ \alpha$. To justify this choice, we do a simple Buckingham Pi analysis and derive the Deborah number as one of the relevant $\Pi$ groups.

One can re-model the system through an experimental lens and pose the problem statement as measuring the average radii $\langle R \rangle$ as a function of time. Intuitively, this might be influenced by system properties like $\epsilon$, $\rho$, $\alpha$, $\xi$. These $4$ quantities along with $\langle R \rangle$, are comprised of the dimensions $M$, $L$ and $T$. Thus $2 \ \Pi$ groups can be made using these variables for every combination of $3$ repeating variables being chosen. Here, we choose $\rho$, $\alpha$, and $\xi$ as are repeating variables.

\begin{align}
\begin{split}
    \Pi_1 & = f(\epsilon, \rho, \alpha, \xi) \\
    & = \epsilon \rho^a \alpha^b \xi^c
\end{split}
\label{eq:pi-eq}
\end{align}

Solving for $a$, $b$, and $c$ so that $\Pi_1$ is dimensionless, we get $\Pi_1 = \epsilon/(\alpha\xi)$ or $De = \xi\alpha/\epsilon$.

\section{\label{secSI:avalanches}Landscape and Non-intermittency at large $\xi$} 
\subsection{\label{subsecSI:avalacnheshighxi}Landscape minima at higher $\xi$}
As discussed in \Cref{sec:avalanches} and \Cref{fig:avalanches} in the main text, the intermittent dynamics and nature of foams can be traced back to their proximity to underlying minima. In \Crefsubfigrange{fig:avalanches}{d}{f}, we see that for higher $\xi$, the system moves further away from energy minima, effectively bypassing them and becoming non-intermittent. However, we observe that for much higher $\xi$, while the system does move higher in $U$ away from the minima floor, the $U_{IS}$ floor also shifts down (see below). For very high values of $\xi = 1.0$, FIRE relaxation consistently finds unjammed minima, indicative of the system being below $\phi_J$.

\begin{figure}[t]
\centering
\begin{captivy}{\includegraphics[scale=1]{\foldername/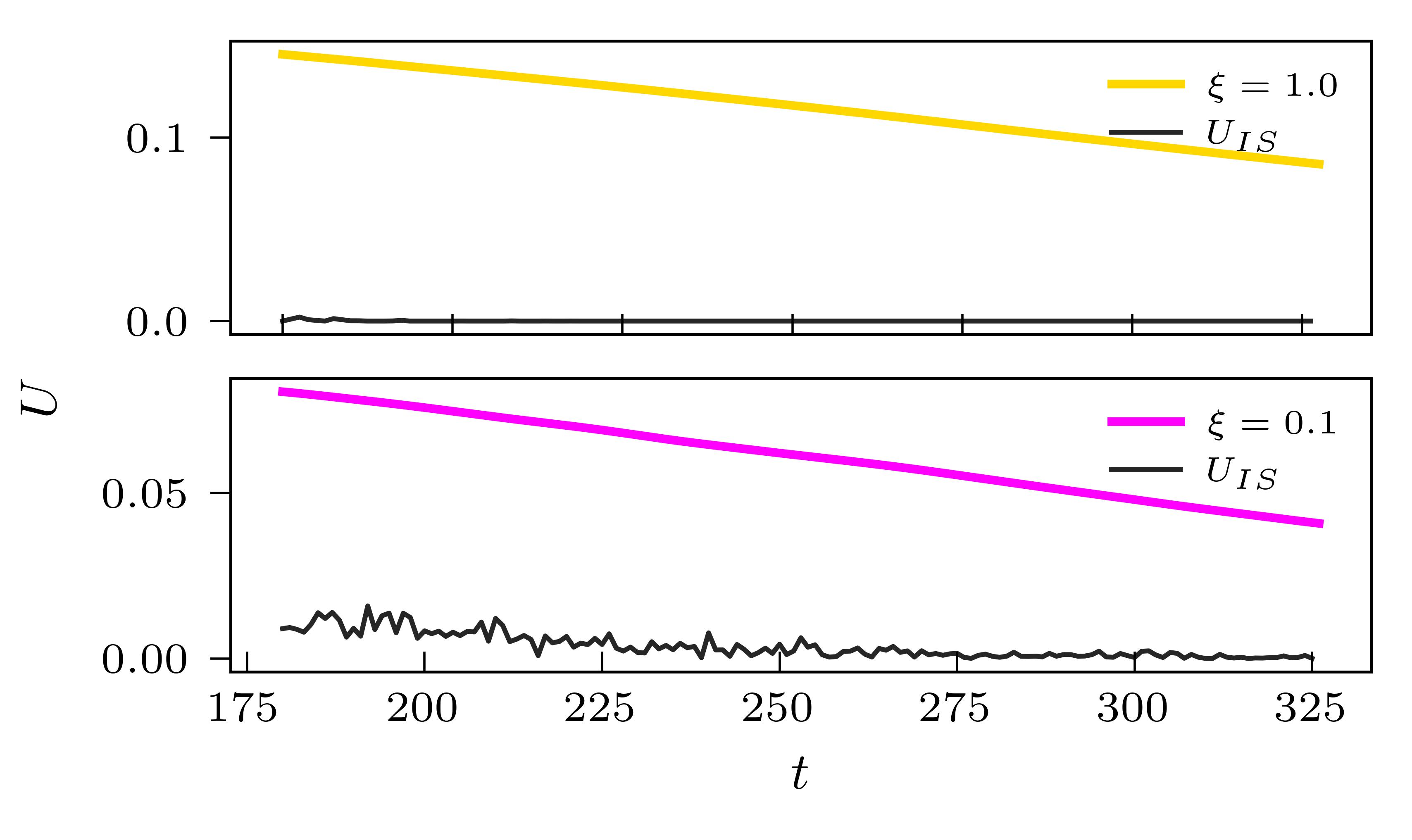}}
    \oversubcaption{-0.05, 0.98}{}{figSI:UISa}
    \oversubcaption{-0.05, 0.48}{}{figSI:UISb}
\end{captivy}
\caption{\textbf{Underlying Landscape}: We perform FIRE relations to find the nearest underlying ISes for simulations at $\xi = 0.1$ and $1$. For the highest $\xi$, the system almost always finds unjammed configurations, while the lower value yields very low-energy ISes.
}
\label{figSI:UIS}
\end{figure}

It should be emphasized that while regions of zero energy open up in the energy landscape at these high $\xi$ values, the ripening systems never explores them. While these chhanges in unexplored energy minima do not affect our conclusions, it does lead us to analyze the radii distribution for any clues regarding changes in the landscape.

\subsection{\label{subsecSI:a2t}Ripening and radii evolution}
As discussed in the main text, ripening mediated evolution of the radii distribution occurs until the system reaches a steady state when the mean normalized radii distribution i.e $P(a_i/\langle a \rangle)$ becomes a constant. This \textit{dynamical scaling} state is observed for all the different $\xi$ simulations. 

\begin{figure}[t!]
\centering
\begin{captivy}{\includegraphics[scale=1]{\foldername/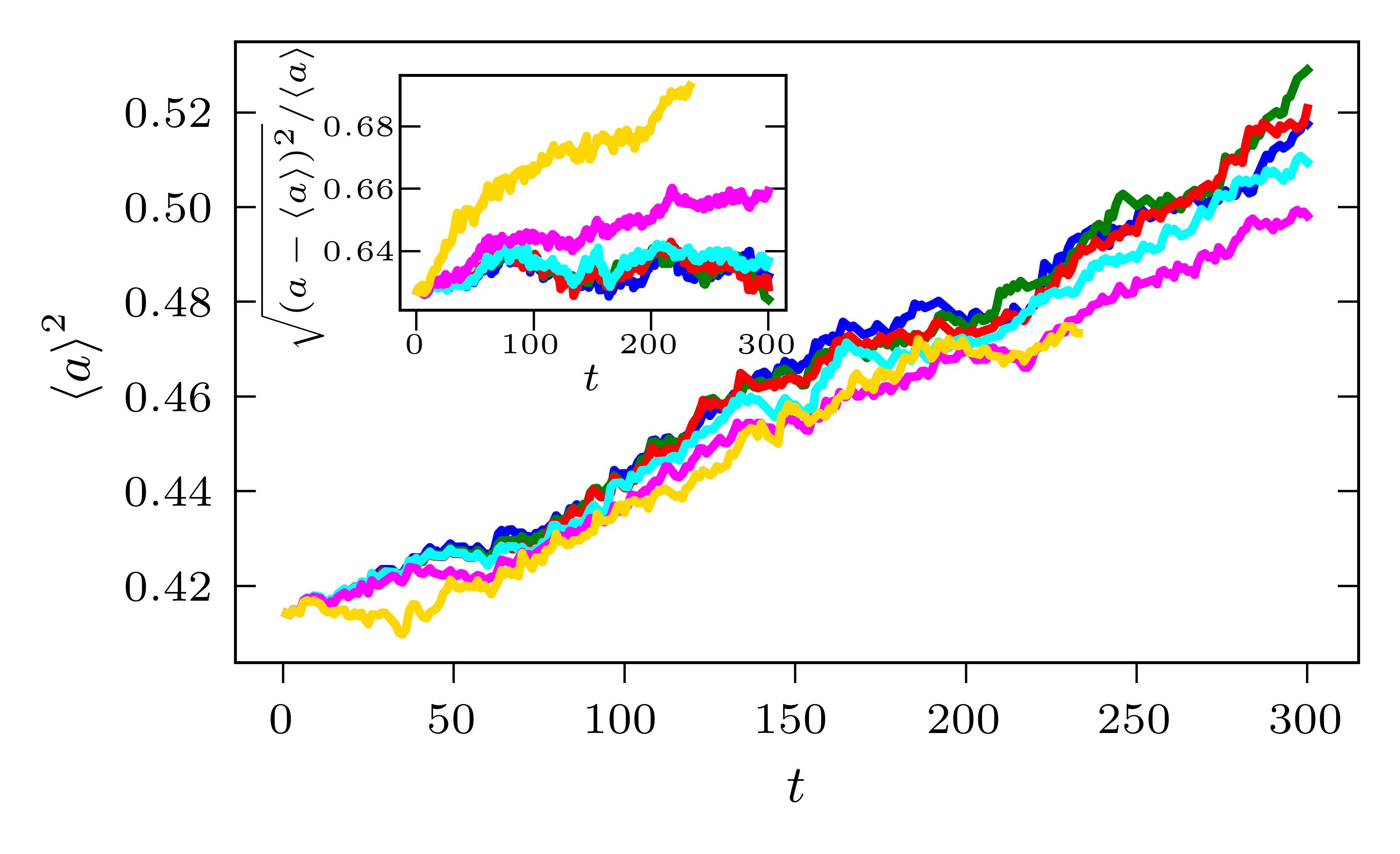}}
\end{captivy}
\caption{\textbf{Radii evolution}: We track the evolution of the bubble radii distribution by measuring two quantities viz. the squared mean (${\langle a \rangle}^2$) and the mean normalized standard deviation ($\sqrt{\langle (a_i - \langle a \rangle)^2 \rangle }/ \langle a \rangle$). While the squared mean evolution is mostly independent of $\xi$ at steady state, the standard deviation increases for large $\xi$, indicative of a more polydisperse foam.
}
\label{figSI:a2t}
\end{figure}

We initialize or begin each simulation with the same initial radii distribution (a Weibull derived from a qausistatic simulation (see \Cref{sec:bubblemodel})). However, we find that at the same volume fraction and ripening rate $\alpha$, these different $\xi$ simulations produce varying steady state distributions. For smaller $\xi$ values, this remains similar to the quasistatic case. For much higher values, the normalized standard deviation $\sqrt{\langle (a_i - \langle a \rangle)^2 \rangle }/ \langle a \rangle$ increases noticeably, while the time evolution of ${\langle a \rangle}^2$ remains similar. This indicates that the bubbles become more polydisperse at higher $\xi$, which is known to lead to higher values of $\phi_J$.  This $\xi$ dependent increase in $\phi_J$ is responsible for the unjamming behavior seen during FIRE relaxation of high viscosity foams. Further, one may note that the time taken to reach steady state i.e to reach a roughly steady distribution, also varies (see \Cref{figSI:a2t}(inset)) and is longer for higher $\xi$ simulations. This, as expected, is due to the initializing bubble size distribution being closer to the quasistatic steady state bubble size distribution than a higher $\xi$ steady state distribution. To consistently stay within dynamical scaling, for all our simulations across various starting initializations, we consider only states beyond $t>180$ for our analysis.

It is worth noting here that the polydispersity of all these ripening systems frequently produces multiple bubbles large enough to span the dimensions of the box. When this leads to multiple overlaps between pairs of bubbles the simulation is ended. Further, an artificial move like a scramble can create multiple overlaps from a configuration that did not have one before scrambling. This restricts our ability to find identical seeds/ starting points across $\xi$s that can be scrambled deep into dynamical scaling, thus requiring us to use different seeds (as in \Cref{fig:reset}).

\section{\label{subsecSI:analyses}Analyses}

\subsection{\label{subsecSI:MSDfractal}MSD and trajectory analysis}
As one of our main analyses, we examine the random motion of the simulation trajectory as a function of lag time or, alternatively, as a function of contour distance in configuration space. The former is an ordinary mean versus lag times of the system-averaged squared displacement ($\Delta r^2$). This quantity is then ensemble averaged over $4$ separate positional initializations (seeds). The average was computed over all pairs of points separated by a given lag time (overcounting). Because of poor statistics at high lag times $\tau$, we determine a maximum lag time for each seed run. This was done such that every run has at least $\sim 100$ independent samples of the MSD data at $\tau_{max}$. The number of independent samples was estimated as the number of non-overlapping lag time intervals (without overcounting) multiplied by a factor of $3$. A common $\tau_{max}$ across the $4$ runs is then chosen as the minimum values across these independent runs. Thus, we effectively get $>100$ independent samples of the data at collective $\tau_{max}$, plotted in \Cref{fig:MSDs}. In the context of rheological analysis, we term this quantity the `strain MSD'.

\begin{figure}[b!]
\centering
\begin{captivy}{\includegraphics[scale=1]{\foldername/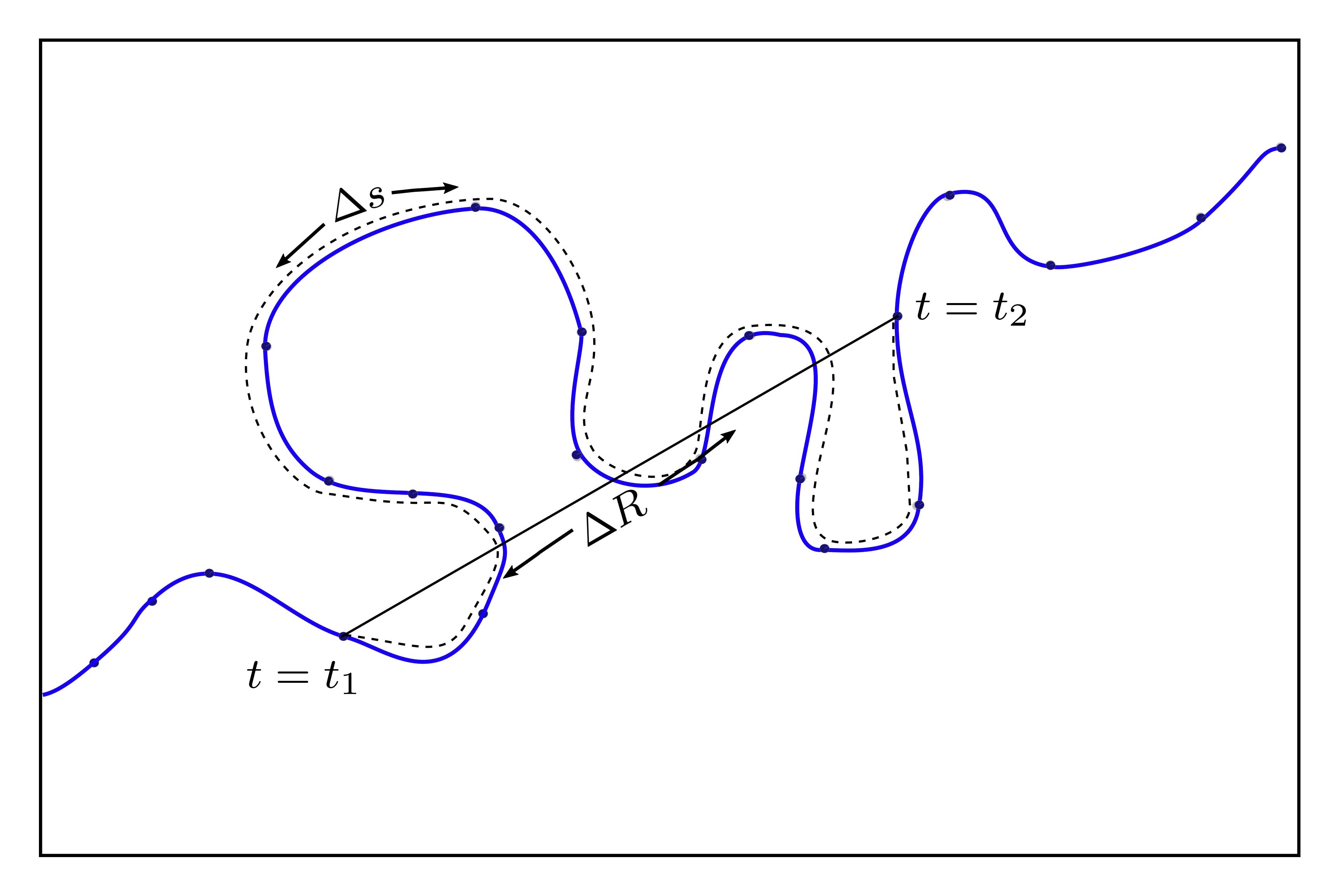}}
\end{captivy}
\caption{\textbf{Trajectory Analysis}: We analyze the trajectories spatially by measuring the Euclidean distance $\Delta R$ and contour distance $\Delta s$ between points sampled (usually at periodic intervals in time) from a ripening simulation.
}
\label{figSI:dr2ds}
\end{figure}

To analyze the high-dimensional path taken by the system, we consider two spatial quantities: the Euclidean displacement $\Delta R$ and contour distance $\Delta s$. This provides a geometrical description of the trajectory (rather than an explicitly dynamical one),  see \Cref{figSI:dr2ds}.  We calculate these between all possible pairs of discrete trajectory points from our simulation. It must be noted that unlike the MSD, which is an average per-bubble quantity, this is $3N$ dimensional quantity. The data collected is then log-binned in $\Delta s$ over fixed bins spanning the whole range of data collected across $4$ independent simulations. Finally, the binned averages of $\Delta R^2$ are averaged over the multiple simulations. We only consider bins that have at least $4$ separate points from each of the $4$ simulations. Figure \Cref{figSI:dr2ds} graphically summarizes the 2 quantities that are considered over these trajectory points in $3N$ dimensions.

\begin{figure}[t!]
\centering
\begin{captivy}{\includegraphics[scale=1]{\foldername/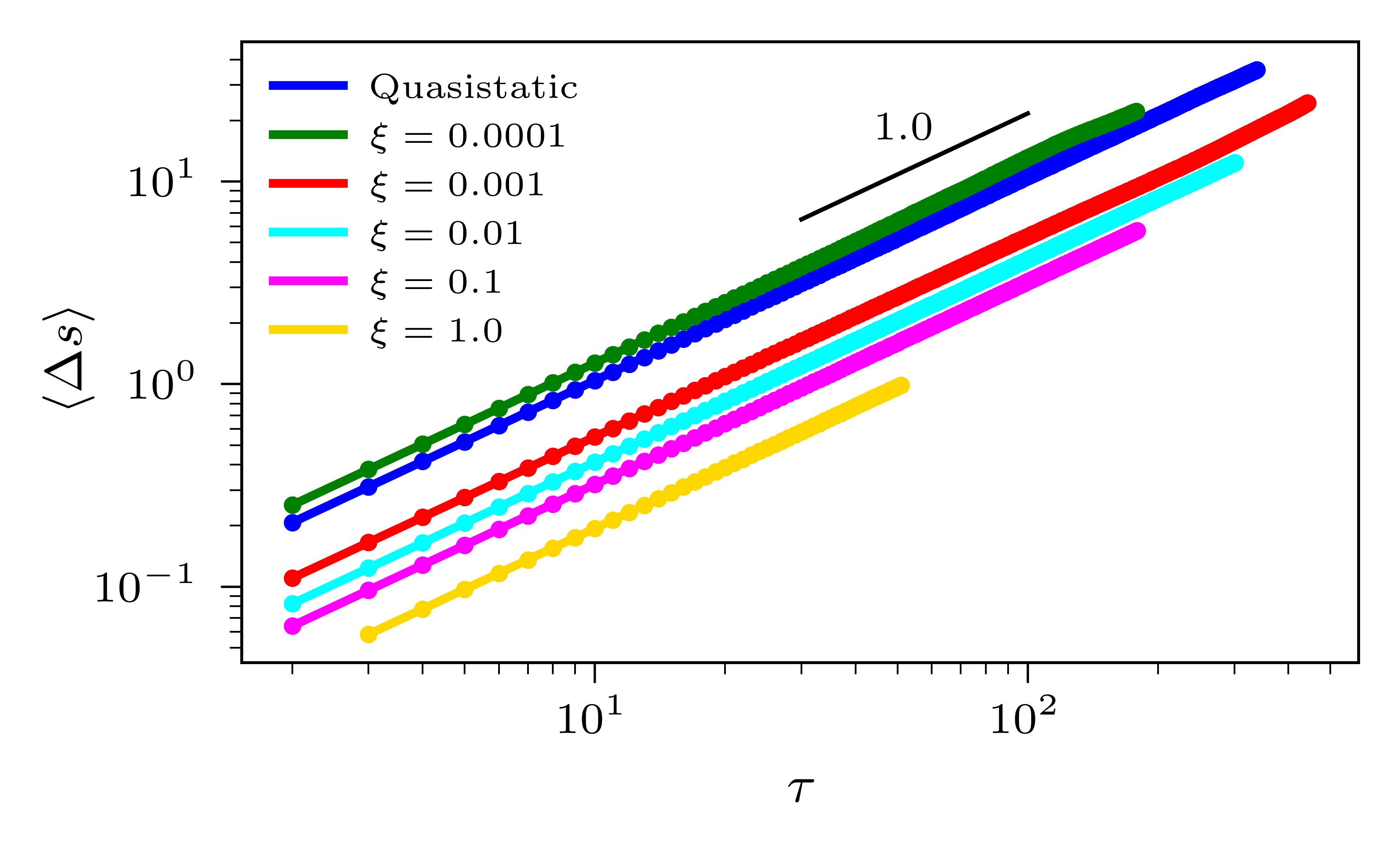}}
\end{captivy}
\caption{\textbf{Trajectory displacements with time}: We measure the ensemble averaged mean $3N$ dimensional displacement of bubbles over various lag times $\tau$. The unit slope indicates the one-to-one relation between the two and the flexibility to switch between both when analyzing rheological and dynamical properties.
}
\label{figSI:dsdt}
\end{figure}

A similar analysis extends to our calculations involving the stress, $\sigma$, and the lag time-based version of this quantity is called the `stress MSD' for our rheological analysis.
For all the above calculations, it must be noted that the various trajectory distances and quantities (like stress) are evaluated over the reduced ensemble of bubbles discussed previously (see \hyperref[sec:Methods]{Methods}). Finally, we analyze the mean contour distance as a function of lag-time below, demonstrating why the slopes analyzed in \Cref{fig:MSDs} and \Cref{fig:stressMSDs} are so similar.

\subsubsection{\label{subsubsecSI:slope} Power-law slopes and eye-guides}

As a first step in analyzing the various time-based and contour distance-based MSDs obtained above, we determine various terminal slopes for these curves (see \Cref{fig:MSDs} and \Cref{fig:stressMSDs}). The lag time-based MSDs are primarily used for calculating rheological properties, thus their terminal slopes are critically important. From previous work \cite{SGMHwang2016}, we know that stress MSDs are more noisy. So, we first analyze the less noisy strain MSD. After our statistics-based truncation, we use the last decade in $\tau$ to determine the power-law scaling using a straight-line fit to the log of the data, where we find $R^2 \ge 0.99$ for all cases presented here. Due to the large fluctuations in the stress MSDs, only the half a decade with the same  lower time limit as the strain MSD yields a straight-line fit with $R^2 \ge 0.99$. The slopes obtained for these curves are $1.36$ and $0.88$, respectively, as shown in \Cref{fig:MSDs}.

For the high-dimensional MSDs (\Cref{fig:stressMSDs}), we follow a similar protocol, finding slopes $1.4$ and $0.9$, respectively, with a $R^2 \ge 0.99$. Similarly, for \Cref{fig:sqrheologya}, we find slopes of $1.35$ and $0.93$, respectively. The lines shown in all these plots correspond to the actual $x$ and $y$ ranges over which the above fits were done. However, for all these MSD-based plots, we show eye-guides of ($1.37$ and $0.9$) instead of the actual fits, comparing and highlighting their similarity. It must also be noted that while this method is quantitative and provides insight into terminal behavior and resulting rheology, the actual values of the slope, especially their second significant digit, are sensitive to the choice of range for the straight-line fit. An uncertainty of at best $0.04$ and $0.08$ in the fit exponents can be assumed for the strain and stress MSDs respectively.

\begin{figure*}[t]
\centering
\begin{captivy}{\includegraphics[scale=1]{\foldername/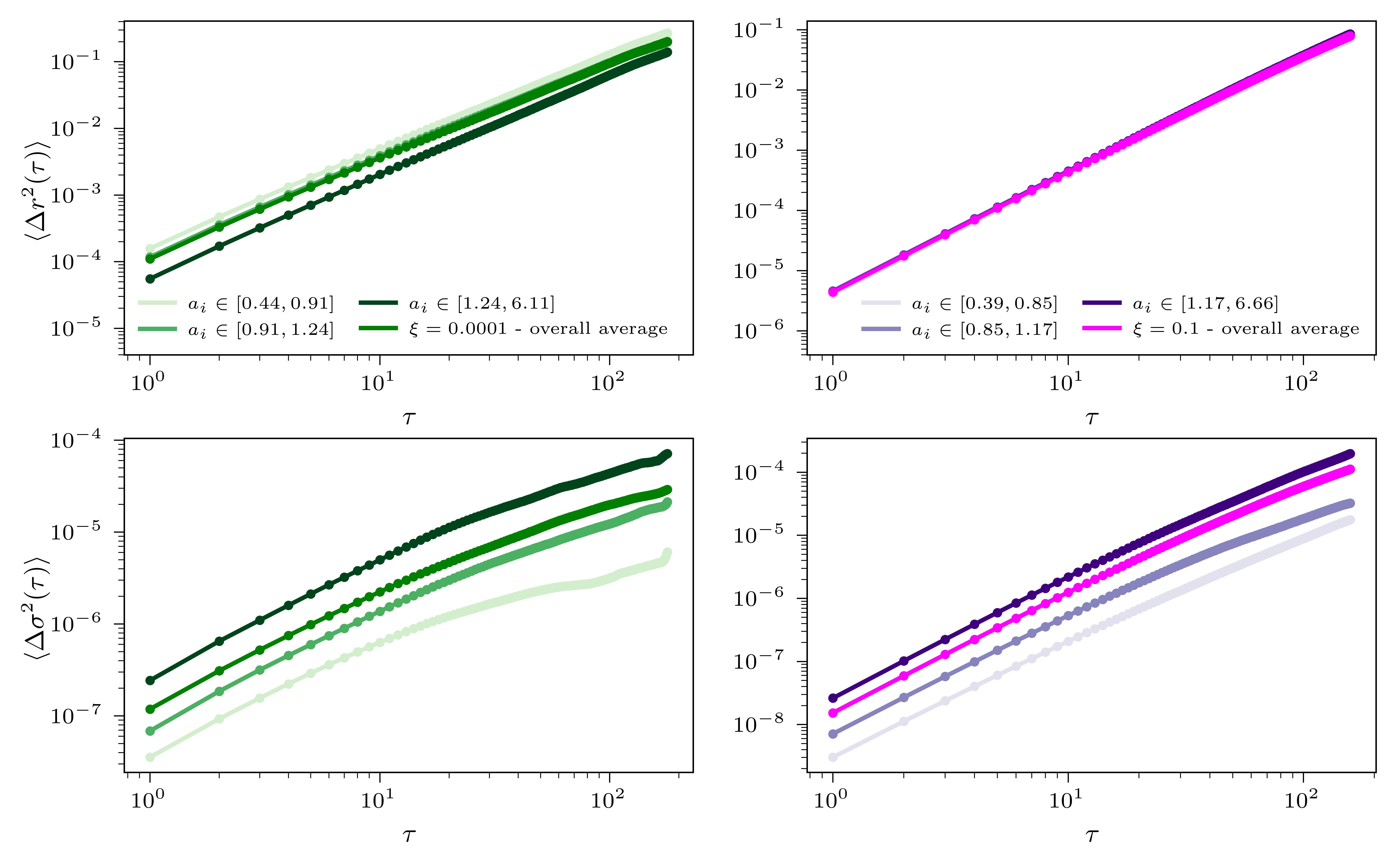}}
    \oversubcaption{0.01, 0.98}{}{figSI:MSDsaia}
    \oversubcaption{0.01, 0.48}{}{figSI:MSDsaib}
    \oversubcaption{0.515, 0.98}{}{figSI:MSDsaic}
    \oversubcaption{0.515, 0.48}{}{figSI:MSDsaid}
\end{captivy}
\caption{\textbf{Sensitivity analysis of strain and stress MSDs on bubble radius}: The bubble ensembles are divided into three tertiles according the their time averaged bubble radii $a_i$, and the MSDs computed for two different values of $\xi$.  This analysis confirms that both the strain MSDS (top row) and stress MSDs (bottom) row of the three tertiles have very similar lag time $\tau$ dependence as the ensemble-averaged MSDs analyzed elsewhere in the text. The amplitude dependence of MSD on $a_i$ is discussed in \Cref{subsubsecSI:msdreadiusdep}
}
\label{figSI:MSDsai}
\end{figure*}

\subsubsection{\label{subsubsecSI:msdreadiusdep} Sensitivity of MSD to Bubble Radius}
The ensemble of bubbles being averaged to compute MSDs is very polydisperse, so we explored the sensitivity of the shape of the MSD on bubble radius. In particular, we wanted to confirm that the terminal power-law exponents reported for the entire ensemble were observed across bubble sizes. To do this, each ensemble was divided into three tertiles based upon each bubble's time averaged radius during the simulation. The results for two different $\xi$ values are shown in \Cref{figSI:MSDsai}. The plots confirm that in an intermittent foam at $\xi = 0.0001$, both strain and stress MSDs do have very similar $\tau$ dependence across tertiles. The smallest bubbles show the largest displacements, as expected in analogy with a Generalized Stokes-Einstein relation. The smallest particles show the lowest stress amplitude, which suggests that stress is  concentrated in larger bubbles. The results for a viscosity dominated $\xi = 0.1$ again finds that both strain and stress MSDs have very similar $\tau$ dependence across tertiles.  As before, smaller bubbles show the smallest stresses, again suggesting stress concentration in larger bubbles. The strain MSDs however are effectively radius independent, which is likely an artifact of the viscous forces in our model being independent of bubble radius, \Cref{eq:damped-equation}. Since the rheology analysis in this study deals only with the frequency dependence of the MSDs, the radius sensitivity of the MSD amplitudes does not affect our conclusions in any way.

\subsection{\label{subsecSI:rheology}Rheology}
\subsubsection{\label{subsubsecSI:rheologymodel}Analytical Derivation}

Here, we provide a derivation for the integral equation \Cref{eq:microrheology}, which was used to compute the viscoelastic moduli for our simulation. We start by noting that the theory of viscoelasticity for linear materials \cite{LinearVEVolterra1909, LinearVEVolterra1913} shows that the creep compliance $J$, can be related to the stress $\sigma$ and strain $\gamma$ as
\begin{align}
\begin{split}
    \int_{-\infty}^{t}{J(t-t^{\prime})\dot{\sigma}(t^{\prime})dt^{\prime} = \gamma(t)}
\end{split}
\label{eq:rheology-derivation-1}
\end{align}

Here we note that $J$, $\sigma$ and $\gamma$ are zero valued for $t \in (-\infty,0)$ and non-negative for $t \in [0,\infty)$. Thus, we can extend the integral limits and perform a subsequent product rule rearrangement based on the behavior of these functions at the limits, yielding
\begin{align}
\begin{split}
    \int_{-\infty}^{\infty}{J(t-t^{\prime})\dot{\sigma}(t^{\prime})dt^{\prime}} & = \gamma(t) \\ \text{\quad \qquad Using the product rule here gives,}\\
    \int_{-\infty}^{\infty}{\dot{J}(t-t^{\prime}){\sigma}(t^{\prime})dt^{\prime}} & = \gamma(t) \\
    \dot{J}(t) \circledast \sigma(t) & = \gamma(t)
\end{split}
\label{eq:rheology-derivation-2}
\end{align}

Taking the Fourier Transform of the convolutional equation above and applying the convolution theorem gives us,

\begin{align}
\begin{split}
   {\widetilde{\dot{J}}} \ \widetilde{{{\sigma}}} & = \ \widetilde{\gamma}, \\
\end{split}
\label{eq:rheology-derivation-3}
\end{align}

\noindent where $\widetilde{x}$ is the Fourier Transform of the function $x(t)$, itself a function of Fourier frequency $\omega$.

While one could numerically integrate \Cref{eq:rheology-derivation-2} or \Cref{eq:rheology-derivation-3} using data from macroscopic measurements, when analyzing large numbers of stochastic trajectories it will be more convenient to convert them to a form based on easily computed mean-squared differences, or MSDs \cite{SGMHwang2016, 2ptMicroRheoCrocker2007}. Thus, we start by considering the squared version of Eq.~\ref{eq:rheology-derivation-3}. Then the Wiener–Khinchin theorem is used to relate the power spectra to the Fourier transforms of the autocorrelation of the same function. Lastly, the autocorrelations can be related algebraically to the mean squared displacement and variances of the functions:

\begin{align}
\begin{split}
   {\widetilde{\dot{J}}} \ {\widetilde{\dot{J}}} 
   & = \frac{{||\widetilde{\gamma}||}^2}{{||\widetilde{\sigma}||}^2} \\
   & = \frac{\widetilde{R_{\gamma \gamma}}}{\widetilde{R_{\sigma \sigma}}} \\
   & = \frac{\widetilde{\langle \gamma^2 \rangle - \langle \Delta \gamma^2 \rangle/ 2}}{\widetilde{\langle \sigma^2 \rangle - \langle \Delta \sigma^2 \rangle /2}} = \frac{\widetilde{\langle \Delta \gamma^2 \rangle}}{\widetilde{\langle \Delta \sigma^2 \rangle }},
\end{split}
\label{eq:rheology-derivation-4}
\end{align}

\noindent where for a given stationary random function $x(t)$,  $||\widetilde{x}(\omega)||^2$ denotes the Fourier power spectral density, $R_{xx}(\tau)$ denotes the temporal autocorrelation function, $\langle x^2 \rangle$ denotes the variance (a constant) and $\langle \Delta x^2(\tau) \rangle$ denotes the mean-squared displacement (MSD). All terms in \Cref{eq:rheology-derivation-4} are functions of frequency $\omega$.

The constants representing the variances only contribute a term at zero frequency after Fourier transformation, $\propto \delta(\omega)$, which can be neglected for finite-frequency measurements, as is standard practice in microrheology and signal processing \cite{OptExpViscModMason1995, G*Mason2000, SwBrownianKrapf2018}. As is conventionally done in microrheology calculations  \cite{G*Mason2000, MicroRheoLubensky2003}, we can quantify continuum strain fluctuations using embedded tracer sphere positions $\mathbf{r}$, via: $ \langle \Delta \gamma^2 \rangle \simeq 3\pi (\langle \Delta \mathbf{r}^2 (\tau_i) \rangle/2) / {\langle a \rangle}^2$. The creep compliance can be related to the dynamic shear modulus using $|J(\omega)||G^*(\omega)| = 1/\omega$ or $|\dot{J}(\omega)||G^*(\omega)| = 1$. Rearranging the equation yields: 

\begin{equation}
    {\lvert{G^*(\omega)}\rvert}^2 \simeq \frac{ \left<a\right>^2 \widetilde{\Delta \sigma^2}(\omega)}{3\pi\widetilde{\Delta\mathbf{r}^2}(\omega)},
\end{equation}

\noindent which corresponds to the formula in the main text, \Cref{eq:microrheology}. Earlier statements of this equation \cite{2ptMicroRheoCrocker2007,SGMHwang2016} contained an incorrect power of $\left<a\right>$ but still yielded the correct frequency dependence.

Formally, the Wiener-Khinchin theorem applies to stationary functions, while the MSD of Brownian motion (and presumably the random processes observed in this study) are \textit{non-stationary} by virtue of being unbounded (as $\tau \rightarrow \infty$).  This subtle issue---effectively when a power spectrum can be reliably computed from a finite segment of a random trajectory---is discussed in detail by Krapf \cite{SwBrownianKrapf2018}.  Briefly, these analyses remain valid when the increments of the random process are stationary. This property is expected to hold for trajectories emerging from a Generalized Langevin Equation with a viscoelastic memory kernel (e.g., fractional Brownian motion \cite{FBMMandelbrot1968} and microrheology \cite{OptExpViscModMason1995}) but not for other processes such as L\'evy flights (where the increments are also non-stationary).

\subsubsection{\label{subsubsecSI:calculatingFTs}Calculating Fourier transforms of experimental MSDs} 

To compute the Fourier transforms of our experimental (log-spaced) MSDs, we follow the approach of Mason \cite{G*Mason2000}, which approximates the MSD as piecewise power law functions of lag time.  The Fourier transform of a power-law defined over the positive axis is  $\widetilde{H(t) t^\alpha} = (i\omega)^{-(\alpha+1)}\Gamma(\alpha+1)$, where $H(t)$ is the Heaviside function. Similarly, we can approximate the Fourier transform of the MSD with the algebraic equation:
\begin{align}
\begin{split}
  \widetilde{MSD}(\omega_i) \approx (1/\omega_i) MSD(\tau_i) \Gamma(1+\alpha(\tau_i)) \big\rvert_{\tau_i = 1/\omega_i}, \\
\end{split}
\label{eq:Mason-approx}
\end{align}

\noindent where the $\{\omega_i\}$ and $\{\tau_i\}$ are discrete values densely spanning the experimental range and $\alpha(\tau) \equiv \partial \log MSD(\tau) / \partial \log \tau$.  This approach is exact when the MSD has a power-law form and has a maximal error $<12\%$ in typically encountered crossover regions \cite{G*Mason2000}. In practice, by numerically fitting the log-slopes and plugging the strain and stress MSD values (computed at $\{\tau_i\}$) into \Cref{eq:Mason-approx} and then \Cref{eq:microrheology}, we obtain the FTs and rheology curves at the corresponding Fourier frequencies $\{\omega_i\}$, as shown in \Cref{fig:MSDs}(inset).

\subsubsection{\label{subsubsecSI:rheologycaluclation} Rheology calculation}

We calculate our rheology curves reported in \Cref{fig:MSDs}, using the \Cref{eq:microrheology}, derived above. However for the purposes of our calculation we ignore the value of $a$ in the formula. Further, to make our log-slope calculations uniform across the range of our data, we sample log-spaced points from our MSDs before calculating the shear modulus. Below, we report $G^*(\omega)$ and $J(\omega)$ (or $\tilde{J}$) curves.

\begin{figure}[h!]
\centering
\begin{captivy}{\includegraphics[scale=1]{\foldername/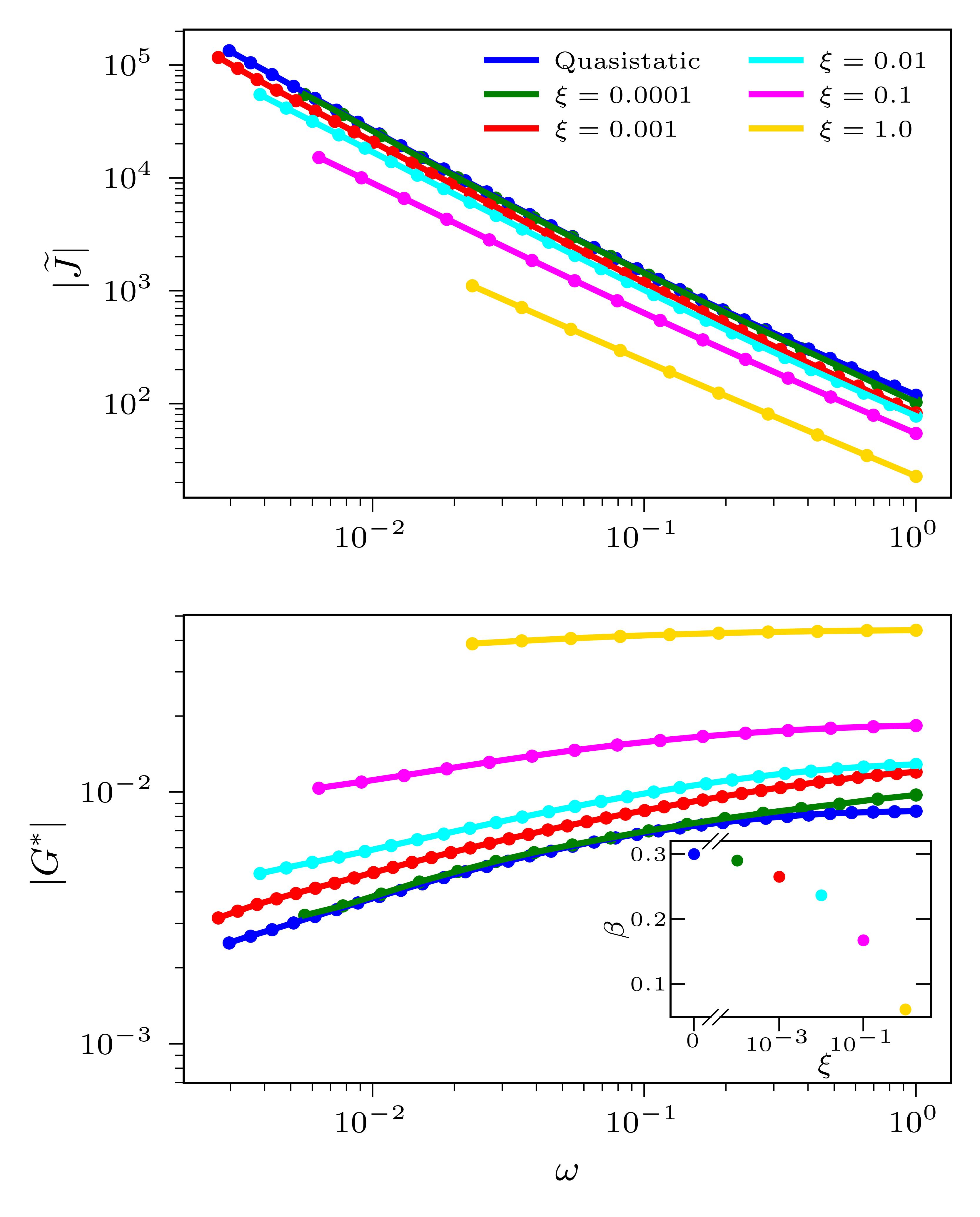}}
    \oversubcaption{0.05, 0.98}{}{figSI:rheologya}
    \oversubcaption{0.05, 0.48}{}{figSI:rheologyb}
\end{captivy}
\caption{{\textbf{Rheology information calculated using the starin and stress MSDs}}:
(\textbf{a}) Fourier transform of the creep compliance is reported above for various $\xi$ values. This can be calculated using \Cref{eq:microrheology} and the relation $|J(\omega)||G^*(\omega)| = 1/\omega$
(\textbf{b}) The complex shear modulus is calculated using the same expression as above. The curves show a clear rollover from power-law behavior to an elastic plateau with an increase in $De$.
(\textbf{inset}) Instantaneous slopes at $\omega \simeq 0.02$ or $\tau \simeq 50$ are reported for the different $De$ shear modulus curves. This broad crossover may explain the varied rheology exponents observed in the experimental literature.}
\label{figSI:rheology}
\end{figure}

\begin{figure}[h!]
\centering
\begin{captivy}{\includegraphics[scale=1]{\foldername/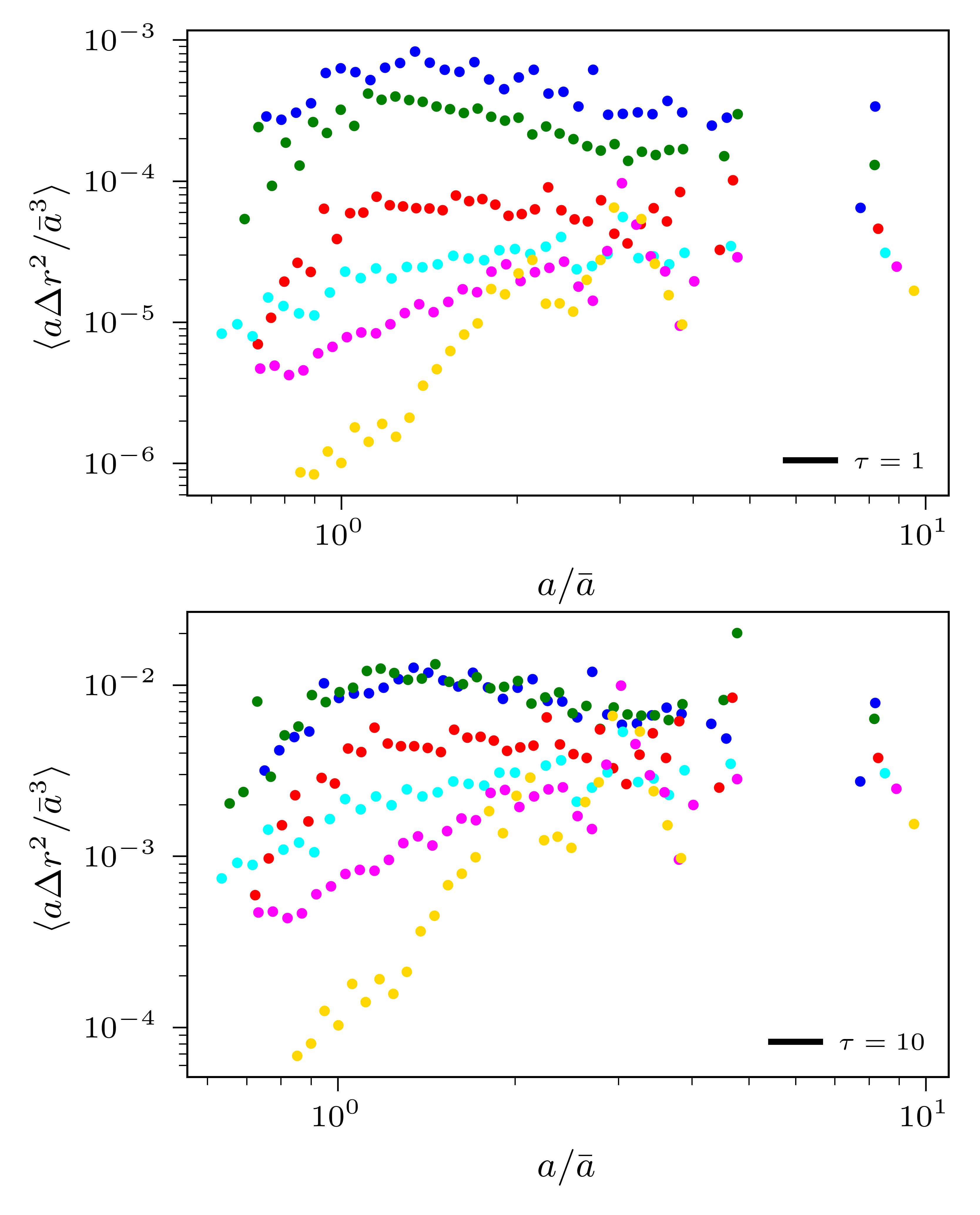}}
    \oversubcaption{0.05, 0.98}{}{figSI:dr2aa}
    \oversubcaption{0.05, 0.48}{}{figSI:dr2ab}
\end{captivy}
\caption{\textbf{Mechanical heterogeneity}: The mean squared displacement of individual bubbles, at $\tau = 1$ (a) and $\tau = 10$ (b), multiplied by their radii, was computed using the same bubble ensemble used for MSDs in the main text, see the \hyperref[sec:Methods]{Methods}. Spherical particles in a homogeneous continuum are expected to display an MSD that varies $\sim 1/a$, which would result in a horizontal line in this plot. For low $De$ the data shows a weakly increasing $a$-dependence whose for small bubbles. AS discussed in \Cref{subsubsecSI:hetergeneity}, the stronger deviations at higher $De$ (and the smallest $a$) are likely an artifact of the viscous force in our model being $a$-independent.
}
\label{figSI:dr2a}
\end{figure}

\subsubsection{\label{subsubsecSI:hetergeneity}Mechanical heterogeneity}
Quantitative microrheology measurements typically assume that the material is a homogeneous continuum at the tracer scale.  In the homogenous case, the observed MSD amplitudes should scale inversely with bubble radius.  Plotting the amplitude of the MSD, at two lag times, versus bubble radius, \Cref{figSI:dr2a}, shows deviations from the expected scaling, in particular for the smallest bubbles. For the lowest $De$, corresponding to landscape-driven foams, the deviations from $1/a$ scaling are quite modest, and may be the effect of dynamical heterogeneity in these systems.  At the highest $De$, the deviations from $1/a$ scaling are more pronounced, closer to showing a constant MSD amplitude with respect to $a$.  As seen previously in \Cref{figSI:MSDsai}, this likely is an artifact related to the magnitude of viscous forces being $a$-independent in \Cref{eq:damped-equation}

\begin{figure*}[t]
\caption{Polydisperse cell list}
\label{alg:polycelllist}
\begin{center}
\fbox{%
\begin{minipage}{0.9\textwidth}
\begin{algorithmic}[1]
\State $l_c \gets \sim \langle a \rangle$ \Comment{Usually chosen around the mean radius}
\State $N_c \gets \texttt{floor}(L_{\texttt{box}} / l_c)$
\State $l_c \gets L_{\texttt{box}} / N_c$
\For{$i \in \{1,2,3, \dots, N\};\ i{+}{+}$}
  \For{$k \in \{1,2,3, \dots, d\};\ k{+}{+}$} \Comment{All $d$ dimensions}
    \State $\texttt{ind}_{i,k} \gets \texttt{floor}(x_{i,k} / l_c)$
  \EndFor
  \State $\texttt{cell-list}_{\texttt{ind}_i} \gets \texttt{append}(i)$ \Comment{Populating the cell list}
\EndFor
\For{$i \in \{1,2,3, \dots, N\};\ i{+}{+}$}
  \State $r_c \gets 2a_i + r_{\text{buffer}}$
  \State $\texttt{ind}_c \gets \texttt{ceil}(r_c / l_c)$
  \For{$l \in \{-\texttt{ind}_c + \texttt{ind}_i, \dots, \texttt{ind}_i + \texttt{ind}_c\};\ l{+}{+}$} \Comment{Loop over all $d$}
    \For{$j \in \texttt{cell-list}_l;\ j{+}{+}$}
      \If{$a_j < a_i$} \Comment{Only neighbors smaller than $i^{\text{th}}$ particle}
        \If{$\texttt{neighbor-criteria}(i, j)$ is \texttt{true}}
          \State $\texttt{neighbor-list}_i \gets \texttt{append}(j)$ \Comment{Non-redundant buffered neighbor list}
        \EndIf
      \EndIf
    \EndFor
  \EndFor
\EndFor
\end{algorithmic}
\end{minipage}%
}
\end{center}
\end{figure*}

\section{\label{secSI:algorithm}Efficient Cell List Algorithm for Polydisperse Systems}
A \textit{cell list} is a commonly used data structure in MD simulations to calculate the neighbors of individual particles. They are used along with \textit{neighbor lists} which keep track of all neighbors that a particle has. Calculating and tracking neighbors is critical to calculating inter-particle forces and propagating dynamics. Since most dispersive force interactions, relevant to soft matter systems have finite cut-offs, it is critical to use computationally efficient ways to calculate nearest neighbors. A standard cell list works by diving the simulation box into cells defined by the cut-off distances for the forces. For forces that only act on contact, the box is divided such that each cell contains about one particle. In both cases, this enables a scenario in which adjacent cells can be checked to determine and populate the neighbor list. For polydisperse systems, choosing the appropriate cell size/dimension is challenging and can lead to computationally inefficient scenarios. We provide a modified cell list algorithm optimized for polydisperse particles with contact forces below.

We start out by building the cell list with cells sized at about the mean bubble radii. In this case, we look for neighbors by evaluating cells further away from those immediately adjacent to accommodate for polydispersity. As an efficient strategy to reduce the number of adjacent cells to check, we maintain a non-redundant neighbor list, with any particle only tracking neighbors which are smaller in size than itself. This effectively limits the number of adjacent cells needed to be checked, while tracking all neighbors non-redundantly. We combine this along with a Verlet list \cite{AlgosFrenkel1996} or a neighbor list with a buffer. Here, additional neighbors or particles beyond contact are tracked, and nearest neighbors are determined by screening this list. This allows for only an occasional need to build the cell list. We implemented a version where net particle displacement counters are tracked over time. When the combined displacements of any two particles are greater than the buffer length, the cell list is called upon to update the non-redundant buffered neighbor list.

In addition to the displacement buffer described above, we account for radii changes in our ripening simulations by tracking the signed magnitude of net particle radii changes. This value is added to the displacements tracked for each particle. Finally, when this composite value for any two particles sums to greater than $r_\text{buffer}$, the cell list and neighbor list are updated. In general, this modified algorithm provides an easy and efficient approach that is computationally faster than other conventional methods \cite{AlgosFrenkel1996, AlgoPlimpton1995, AlgoTorquato2005}.

\bibliography{references}

\end{document}